\documentclass[a4paper]{article}

	\usepackage{amssymb}
	\usepackage{amsmath}
	\usepackage{bm}
	\usepackage{graphicx}
	\usepackage{subfigure}
	\usepackage{hyperref}
	\usepackage{color}
	
	\usepackage{booktabs}
	\usepackage{array}
	\usepackage{enumitem}
	\usepackage{cite}
	\usepackage[affil-it,auth-lg]{authblk}
	\usepackage{eso-pic}

\newcommand{\LA}{\left\langle}
\newcommand{\RA}{\right\rangle}

\begin{document}

\title{Statistical Mechanics of Gravitational Systems with Regular Orbits: \\Rigid Body Model of Vector Resonant Relaxation}
\author[1]{Zacharias Roupas}
\affil[1]{Centre for Theoretical Physics, The British University in Egypt, Sherouk City 11837, Cairo, Egypt} 

\date{\vspace{-5ex}}

\maketitle

\begin{abstract}
	I consider a self-gravitating, N-body system assuming that the N constituents follow regular orbits about the center of mass of the cluster, where a central massive object may be present. I calculate the average  over a characteristic timescale of the full, N-body Hamiltonian including all kinetic and potential energy terms. The resulting effective system allows for the identification of the orbital planes with N rigid, disk-shaped tops, that can rotate about their fixed common centre and are subject to mutual gravitational torques. The time-averaging imposes boundaries on the canonical generalized momenta of the resulting canonical phase space. I investigate the statistical mechanics induced by the effective Hamiltonian on this bounded phase space and calculate the thermal equilibrium states. These are a result of the relaxation of spins' directions, identified with orbital planes' orientations, which is called vector resonant relaxation. I calculate the dependence of spins' angular velocity dispersion on temperature and calculate the velocity distribution functions. I argue that the range of validity of the gravitational phase transitions, identified in the special case of zero kinetic term by Roupas, Kocsis \& Tremaine, is expanded to non-zero values of the ratio of masses between the cluster of N-bodies and the central massive object. The relevance with astrophysics is discussed focusing on stellar clusters. The same analysis performed on an unbounded phase space accounts for continuous rigid tops.
\end{abstract}
	
\section{Introduction}

The statistical mechanics of self-gravitating systems \cite{Padmanabhan_1990} belongs to the general research field called statistical mechanics of long-range interacting systems \cite{Campa_2014} and has been enjoying growing attention \cite{1986MNRAS.219..285T,2000ApJ...531..739N,PhysRevE.68.016108,2001WSLNP..66.....G,dvs1,dvs2,2002PhRvL..89c1101V,2005MNRAS.361..385A,2006IJMPB..20.3113C,Axenides_2012PhRvD..86j4005A,2014JPhA...47C2001T,2012PhRvE..85f1105C,2015PhRvD..91f3531C,2015ApJ...807..157T,PhysRevE.99.022108,2019PhRvL.123b1103T}. A long-range interacting system is defined \cite{CAMPA200957} as the one with magnitude of  inter-particle potential energy bounded by $r^{-a}$, with $a\leq d$, where $d$ denotes the dimensionality of the system and $r$ the distance between two particles.
Despite the advances, the theoretical foundation of statistical mechanics of self-gravitating systems faces difficulties and it remains an open, flourishing subject, not as yet fully understood. The fundamental difficulties were first realized in 1962, when Antonov showed that there exists no global entropy maximum state of the Newtonian self-gravitating gas \cite{antonov}. Nevertheless, it was realized that equilibrium states, which are stable at certain timescales and corresponding to local entropy maxima, can exist \cite{LyndenBell:1966bi,1968MNRAS.138..495L}, despite the presence of the gravothermal instability \cite{1980MNRAS.191..483L}. More recently, I generalized the later to the relativistic regime and identified its relativistic counterpart \cite{Roupas_CQG_RGI_2015,Roupas_2019}, showing that the gravothermal instability has two aspects, a low-energy and a high-energy one. 
An integrated part of the gravothermal instability is inequivalence of ensembles \cite{PhysRevLett.87.030601,2005JSP...118.1073B} and stable equilibrium states with negative specific heat \cite{1968MNRAS.138..495L,Velazquez_2016,Roupas_2019}. These issues are also related to the problem of defining a canonical ensemble in long-range interacting systems, tackled by several authors \cite{2001WSLNP..66.....G,2005JSP...118.1073B,2013PhRvE..87d2110D}. 

Here, I focus on the statistical mechanics of a long-range interacting system, motivated by astrophysics. Systems like stellar clusters, that may host a central massive black hole, or planetary systems, contain bodies that follow regular orbits around the cluster's center of gravity. The orbits are regular for most of the bodies inside the radius of influence of a central massive object. This might also be the case for nearly-spherical clusters not hosting a central massive object, like Globular Clusters, at least significantly close to the center (certainly inside the radius containing half the mass of the cluster), though at the outer parts of such clusters most orbits may be irregular (chaotic) \cite{1999CeMDA..73..159C}.

For such type of systems Rauch \& Tremaine \cite{Rauch+Tremaine_1996}  found that the statistical relaxation of vectors of in-plane orbital angular momentum proceeds towards thermal equilirbium faster and independently than the rest degrees of freedom in a process called Resonant Relaxation. This is a type of internal thermal equilibrium achieved and applied at limited timescales. The orbital angular momentum's vectors' directions (the orbital planes' orientations) relax in several, realistic circumstances independently from their magnitudes, in which case the process is called Vector Resonant Relaxation (VRR). The relaxation of orbital angular momentum's magnitudes is called Scalar Resonant Relaxation. Resonant Relaxation has been studied in astrophysical settings \cite{Hopman_2006,Kocsis+Tremaine_2011,Alexander_2017,2018PhRvL.121j1101S,Hamers_etal_2018} especially with numerical simulations \cite{Rauch+Ingalls_1998,Gurkan+Hopman_2007,Eilon_2009,Meiron_2018arXiv180607894M}, but also on a kinetic theory basis \cite{Sridhar+Touma_I,Sridhar+Touma_II,Sridhar+Touma_III,2018ApJ...860L..23B,2018arXiv181207053F}. 

The method of time-averaging of gravitational orbits and their approximation with rigid wires was introduced  by Gauss and has been extensively used in planetary dynamics \cite{Touma-Tremaine-Kazandjian_2009}. In Ref. \cite{2015MNRAS.448.3265K}, the time-averaging was applied in a VRR system without any reference to a kinetic energy term. A dynamics of non-canonical variables (the components of orbital planes' direction vector) satisfying the $SO(3)$ algebra on a non-canonical phase space is induced by solely the effective potential energy term of VRR. For this dynamics, Roupas, Kocsis \& Tremaine \cite{2017ApJ...842...90R} identified gravitational phase transitions in VRR. They calculated the spacial distribution of orbital planes' orientation vectors at thermodynamic equilibrium. 

In this work, I will again apply the time-averaging method over the apsidal precession's time-scale, but now on the full $N$-body Hamiltonian, with all kinetic terms consistently included. The resulting ``rigid-body decomposition'' of the effective energy accounts for three terms determining the evolution; namely, a rotational, normal kinetic term accounting for the orbital planes' precession and nutation, a spin kinetic term accounting for the in-plane rotation and the gravitational interaction term at quadrapole and higher order. This effective Hamiltonian describes rigid, disk-shaped, spinning tops allowed to rotate about any of their diameters crossing the common fixed centre, in direct analogy with rigid body dynamics \cite{Goldstein_2002} Torques on each disk develop due to mutual gravitational attraction. The general dynamical equations of motion of VRR are calculated in the rigid-body decomposition. They naturally induce new physical parameters, which connect the physical properties of the effective system (rigid annular disks) with these of the implicit system (orbiting point masses). These parameters are the moments of inertia and spin magnitudes of the effective rigid disks. They are connected with the averaging time-scale and the ratio $\varepsilon$ of the mass of the cluster to that of the central object. The gravitational couplings mediate the two views --implicit and effective-- of the system and allow for such relations to emerge. The aforementioned $SO(3)$ evolution induced by a zero kinetic term turns out to be  the approximation of the special limit $\varepsilon=0$ at zeroth order. More importantly, the identified relations between properties of the implicit and effective systems allow for the generalization of the dynamics and the validation and further generalization of the gravitational phase transitions in the cases that the clusters' mass is comparable to that of the central massive object. I specify the dynamical conditions for which such generalization may be valid. Last, but not least, I calculate the dependence of the dispersion of disks' precession and nutation on temperature. It depends on $\varepsilon$ and moments of inertia in a non-trivial way. Due to the later dependence, it is possible that different families of bodies acquire different dispersions, even at orders of magnitude.

Note that VRR resembles mathematically in certain aspects the Hamiltonian mean-field model (HMF) \cite{1995PhRvE..52.2361A,2005EPJB...46...61C} and the interested reader might find instructive the analogy.

In the next section \ref{sec:Dyn} I time-average the self-gravitating $N$-body Hamiltonian, demonstrate the equations of motion that emerge and calculate the boundaries of the effective, canonical phase space. In section \ref{sec:Stat} I develop in detail the statistical mechanics of the system. I formally define the microcanonical, the canonical and the Gibbs-canonical ensembles and consider a thermodynamic limit. In section \ref{sec:ineq_ens}, I discuss the inequivalence of ensmbles. In section \ref{sec:GPT} I review, validate and generalize the VRR gravitational phase transitions. In section \ref{sec:K_VRR} I inspect the kinetic energy term and calculate the dependence of the velocity dispersion on temperature. In section \ref{sec:RB} I briefly modify the analysis to account for continuous rigid bodies. In the final section \ref{sec:conclusions} I discuss the results. 

\section{Dynamics of Self-Gravitating Systems with regular Orbits}\label{sec:Dyn}

\subsection{The N-body Energy}

Consider a self-gravitating $N$-body system, consisted of $N$ bodies with masses $m_i$, $i = 1,\ldots ,N$, and a massive body with mass 
\begin{equation}
	M_\bullet \gg m_i,
\end{equation}
which lies fixed in the centre of mass.
The energy of this self-gravitating $N$-body system may be written as
\begin{equation}
	E_\text{N-body} = \sum_i \left\lbrace\frac{1}{2}m_i |\bm{v}_{i}(t)|^2  - G\frac{m_i M_\bullet}{\left|\bm{r}_i(t)\right|} \right\rbrace  - \sum_{i>j} \left\lbrace G\frac{m_i m_j}{\left|\bm{r}_j(t)-\bm{r}_j(t)\right|}\right\rbrace.
\end{equation}
I denote $\bm{v}_i$, $\bm{r}_i$, $m_i$, the velocity and position vector of each body with respect to the center of mass, and the mass of each body, respectively. In this frame of reference the dipole and all other odd terms of $l$ vanish in a multipole expansion \cite{1998clel.book.....J}
\begin{equation}\label{eq:mul_exp}
	\frac{1}{\bm{r}_i - \bm{r}_j} = \sum_{l=0}^{\infty}\frac{\left(\min{\{|\bm{r}_i|,|\bm{r}_j|\}}\right)^l}{\left(\max{\{|\bm{r}_i|,|\bm{r}_j|\}}\right)^{l+1}}P_l (\hat{\bm{r}}_i\cdot \hat{\bm{r}}_j),
\end{equation}
where $P_l$ denotes the Legendre polynomials (for a relevant, though different, approach regarding the expansion of potential energy see also \cite{PhysRevE.61.6270}). The monopole $l=0$ term contributes the energy $U_0 = -GM_{\star,i}/r_i$, where $M_{\star,i} = M_\star(r<r_i)$ denotes the cluster's mass enclosed within $r_i$. The potential energy may be decomposed to the sum of $U_0$ and the quantity $U_\text{Q}$, which itself denotes the sum between quadrapole $l=2$ and higher order terms (but also monopole contributions on mass $m_i$ from $r>r_i$, which will be later neglected due to symmetry considerations)
\begin{equation}\label{eq:U_Q}
	U \equiv - \sum_{i>j} G\frac{m_i m_j}{\left|\bm{r}_j-\bm{r}_j\right|} =
	-\sum_i G\frac{m_i M_{\star,i}}{|\bm{r}_i|} + U_\text{Q} .
\end{equation}
Therefore, finally the total energy of the system may be decomposed as
\begin{equation}\label{eq:E_Nbody_dec}
	E_\text{N-body} = E_\text{K}(t) + U_\text{M}(t) + U_\text{Q}(t),
\end{equation}
where
\begin{align}
	E_\text{K}(t) &\equiv \sum_i \frac{1}{2}m_i |\bm{v}_{i}(t)|^2  ,\\
	U_\text{M}(t) &\equiv - \sum_i  G\frac{m_i M_i}{\left|\bm{r}_i(t)\right|} ,
\end{align}
and 
\begin{equation}
	M_i \equiv M_\bullet + M_{\star,i} = M_\bullet + M_{\star}(r<r_i).
\end{equation}
The $N$-body decomposition (\ref{eq:E_Nbody_dec}) is the one, most useful for our purposes. The monopole term $U_\text{M}$ accounts for the binding of each body $m_i$ with the cluster. If the quadrapole and higher contributions were to be neglected all bodies would follow planar Keplerian orbits, not interacting mutually. Our analysis will account for the case that the quadrapole and higher contributions are such, that at certain timescales they disturb the orbital planes' orientation, but not the in-plane elements of each orbit.

The current analysis applies to the case
\begin{equation}\label{eq:M_cond_gtr}
	M_\bullet \gg mN.
\end{equation}
but also it may be good approximation in the case (which induces different characteristic timescales)
\begin{equation}\label{eq:M_cond}
	M_\bullet \gtrsim mN.
\end{equation}
It is relevant even in the absence of a central massive object
\begin{equation}\label{eq:M_cond_zero}
	M_\bullet = 0,
\end{equation}
since it has been shown in Ref. \cite{Meiron_2018arXiv180607894M} that vector resonant relaxation and the rigid body approximation apply under certain conditions in Globular Clusters not hosting an intermediate massive black hole.
Astrophysically, the condition (\ref{eq:M_cond_gtr}) may be expected to hold at certain parts of a star cluster hosting a supermassive black hole, called a Nuclear Star Cluster in the centre of certain galaxies \cite{Hopman_2006,annurev-astro-091916-055306}. The condition (\ref{eq:M_cond}) may be expected to be relevant among others with core collapsed Globular Clusters that host an intermediate massive black hole surrounded by a black hole subcluster \cite{2017Natur.542..203K,2018MNRAS.478.1844A}.

\subsection{Timescales}\label{sec:timescales}

I assume that the bodies follow regular orbits. The orientation of each ellipse on the orbital plane (the line of apsis) fluctuates (apsidal precession) due to the rest of the bodies at timescales 
\begin{equation}\label{eq:t_aps}
	t_{\text{aps}} \sim \frac{M}{M_{\star}} t_\text{orb},
\end{equation}	
with $M = M_\bullet + M_\star$. For condition (\ref{eq:M_cond}) it is of about the same order, though longer, than the orbital period 
$
	t_{\text{orb}} \sim \sqrt{a^3/G M}
$,
where $a$ is the semi-major axis. The orbital angular momentum directions (orientation of orbital planes) reaches statistical equilibrium at the vector resonant relaxation timescale \cite{Rauch+Tremaine_1996}
\begin{equation}\label{eq:t_VRR}
	t_\text{VRR} \sim \sqrt{N} t_\text{aps}.
\end{equation}
Note that the precession and nutation of orbital planes (not the apsidal precession) fluctuates in a nearly spherical potential, satisfying condition (\ref{eq:M_cond}), at the same timescale $
	t_\perp \sim t_\text{VRR}
$
\cite{Rauch+Tremaine_1996}.
I will average over a timescale $t_\text{av}$ with respect to which the orbital, binding, energy (and therefore the semi-major axis $a_i$) and angular momentum magnitude (and therefore the eccentricities $e_i$) are adiabatic invariants  
\begin{equation}
	t_\text{orb} < t_\text{av} \ll t_\text{VRR} \ll t_\text{SR},\; t_\text{2b}
\end{equation}
where $t_\text{SR} = (M/m)t_\text{orb}$ denotes the characteristic relaxation timescale of angular momentum magnitudes, called scalar resonant relaxation, and $t_\text{2b} = (M^2/Nm^2)t_\text{orb}$ that of non-resonant two-body relaxation \cite{Rauch+Tremaine_1996}. For condition (\ref{eq:M_cond}) we have $t_\text{SR} \sim t_\text{2b}$ and for condition (\ref{eq:M_cond_gtr}) we have $t_\text{SR} \ll t_\text{2b}$.
The $t_\text{av}$ can be chosen to be of the order of the apsidal precession timescale
\begin{equation}
t_\text{av} \sim t_\text{aps}
\end{equation} 
For example it can be the maximum apsidal precession period at the outskirts of the cluster $r=R$, namely 
$
	t_{\text{av}} \sim (M/M_\star)\sqrt{R^3/G M}
$.

The orbits can therefore be averaged over a sufficient $t_\text{av}$ keeping the semi-major axis and eccentricity of each orbit fixed, and also assuming that the orbital plane's orientation is not significantly varied due to gravitational torques in this timescale. For condition (\ref{eq:M_cond_gtr}) it is $t_\text{orb} \ll t_\text{aps}$ and the result of averaging is a non-homogeneous annular disk. For conditions (\ref{eq:M_cond}), (\ref{eq:M_cond_zero}) it is $t_\text{orb} \sim t_\text{aps}$ and the result of time-averaging is a rosette type of shape which is non-closing. Averaging over a few apsidal precession periods (for $i$ star with $a_i < R$) the resulting shape resembles a rigid, annular disk, or a ring for small eccentricities. In Appendix \ref{app:Keplerian} I discuss Keplerian orbits and orbits subject to apsidal precession and I further discuss averaging over time in Appendix \ref{app:av}.

\subsection{Rigid-Body Decomposition}

The velocity $\bm{v}_i$ may be decomposed to a planar velocity $\bm{v}_{\parallel,i}$, that is the velocity component parallel to the orbital plane, and a normal velocity component $\bm{v}_{\perp,i}$ that is orthogonal to the orbital plane. The $N$-body decomposition may be averaged over time as
\begin{equation}\label{eq:E_VR_dec-def}
	E \equiv K_\perp + K_\parallel + U_\text{s} + U_\text{VR},
\end{equation}
where ``VR'' stands for vector resonant relaxation and
\begin{align}
\label{eq:kinetic_normal_def}
	K_\perp &\equiv \sum_i \LA \frac{1}{2}m_i v_{\perp,i}^2 \RA_{t_\text{av}}, \\
	\label{eq:kinetic_planar_def}
	K_\parallel &\equiv \sum_i \LA \frac{1}{2}m_i v_{\parallel,i}^2 \RA_{t_\text{av}}, \\
	\label{eq:U_B_def}
	U_\text{b} &\equiv - \sum_i \LA G\frac{m_i M_i}{r_i}\RA_{t_\text{av}}, \\
\label{eq:U_VR_def}
	U_{\text{VR}} &\equiv \LA U_\text{Q} \RA_{t_\text{av}},
\end{align}
where $\LA \bullet\RA_{t_\text{av}}$ denotes the average over a timescale $t_\text{av}$. The normal kinetic term $K_\perp$ represents the kinetic rotational energy about any diameter of the orbital planes (energy of precession and nutation of orbital planes), the spin kinetic term $K_\parallel$ represents the in-plane kinetic rotational energy (energy of Keplerian rotation and apsidal precession), the binary term $U_\text{b}$ represents the gravitational interaction at monopole order of each body with the cluster (interaction energy of binaries, each composed of the two masses $m_i$ and $M_i$), the VR-interaction term $U_\text{VR}$ represents the gravitational interaction between orbits at quadrapole (and higher) order.

I show in Appendix \ref{app:av} that the first term (\ref{eq:kinetic_normal_def}) may be written as
\begin{equation}\label{eq:K_normal}
	K_\perp (t) = \sum_i \frac{1}{2} I_i \omega_{\perp,i}(t)^2,
\end{equation}
where $I_i$ is the constant moment of inertia of the $i$ disk (all are non-homogeneous) given in (\ref{eq:I_inertia_app}) for nearly-Keplerian orbits.
The normal angular velocity $\bm{\omega}_{\perp,i}(t)$ depends on time, because each orbit is subject to torques from the rest of the orbits due to $U_\text{Q}$. The primary effect of gravitational potential during the timescales considered is the evolution of $\bm{\omega}_{\perp,i}(t)$, which describes the precession and nutation of the orbital planes. The system relaxes towards thermal equilibrium, because of the exchange of potential and kinetic energy, between the disks. The arithmetic parameter $(1/2)$ highlights the analogy with rigid body dynamics \cite{Goldstein_2002} and is introduced due to the perpendicular axis theorem applied to an annular disk. The energy $K_\perp$ is the rotational kinetic energy of the disks, due to the rotation of each one about any of its diameters.

I assume that the second term (\ref{eq:kinetic_planar_def}) may be written as
\begin{equation}\label{eq:K_planar}
	K_\parallel = \sum_i I_i \omega_{\parallel,i}^2 = \text{const.}.
\end{equation}
It is constant in the timescales considered. The moment of inertia $I_i$ in expression (\ref{eq:K_planar}) is the same with that of (\ref{eq:K_normal}). In this form, $K_\parallel$ represents the spinning kinetic energy of the annular disks, described by the normal kinetic term, where `spinning' refers to the rotation of a disk about the primary axis perpendicular to the plane of the disk. The angular velocity $\omega_{\parallel,i}$ is given for nearly-Keplerian orbits in Eq. (\ref{eq:om_par_app}).

The binary term $U_\text{b}$ adds only a constant negative shift to the Hamiltonian in the timescales considered. It is for nearly-Keplerian orbits $U_\text{b} = -\sum Gm_iM_i/a_i$ (see Appendix (\ref{app:av}). The role of this monopole term $U_\text{b}$ is to keep the bodies $m_i$ bound to the cluster and will be neglected from the VR-Hamiltonian as a constant shift to the potential energy. Its role is similar to the role of the electromagnetic binding energy which binds the molecules of a continuous rigid body and allows us to identify it as such. 

Now, regarding the interaction term (\ref{eq:U_VR}) of the rigid-body decomposition it can be written, using the multipole expansion (\ref{eq:mul_exp}), as
\begin{equation}\label{eq:U_VR}
	U_{\text{VR}} = -\frac{1}{2} \sum_{i\neq j,l} \mathcal{J}_{ij,l} P_l(\bm{n}_i(t)\cdot\bm{n}_j(t))
\end{equation}
with $\mathcal{J}_{ij,l} = \mathcal{J}_{ij,l}(m_i,a_i,e_i,m_j,a_j,e_j)$. This term expresses the gravitational interaction, at quadrapole and higher order between disks with a common fixed centre. The unit normal vectors $\bm{n}_i$ are parallel to the orbital angular momentum and perpendicular to the orbital plane of the disk $i$. By definition
\begin{equation}
	\dot{\bm{n}}_i = \bm{\omega}_{\perp,i}.
\end{equation} 
Regarding the coupling constants $\mathcal{J}_{ij,l}$, an expression is given for example in \cite{2015MNRAS.448.3265K} for condition (\ref{eq:M_cond_gtr}). Here we do not need a specific expression.

To conclude, I consider the following rigid-body decomposition of the time-averaged energy
\begin{equation}\label{eq:VR_dec}
	E = K_\perp + K_\parallel + U_\text{VR} \equiv \sum_i \frac{1}{2} I_i \omega_{\perp,i}(t)^2 + \sum_i I_i \omega_{\parallel,i}^2 -\frac{1}{2} \sum_{i\neq j,l} \mathcal{J}_{ij,l} P_l(\bm{n}_i(t)\cdot\bm{n}_j(t)) .
\end{equation}

\subsection{The Lagrangian}

Consider a spherical coordinate system with its centre fixed at the centre of mass of the cluster and denote with $(\theta_i,\phi_i)$ the positions of the unit normals 
\begin{equation}
	\bm{n}_i = (\sin\theta_i \cos\phi_i,\sin\theta_i\sin\phi_i,\cos\theta_i),
\end{equation}
and $\psi_i$ the additional Euler angle, which describes the spin of the disk $i$, that is the rotation about the primary axis which crosses disk's center perpendicularly to the disk. It is by straightforward geometrical considerations \cite{Goldstein_2002}
\begin{equation}\label{eq:omega_def}
	\bm{\omega}_\perp = \dot{\theta}_i \hat{\bm{e}}_\theta + \sin\theta_i \dot{\phi} \hat{\bm{e}}_\phi,\quad
	\bm{\omega}_\parallel = \left(\dot{\psi}_i + \cos \theta_i \dot{\phi}_i\right) \hat{\bm{e}}_\psi .	
\end{equation}
Thus, the Lagrangian of the system is written as
\begin{equation}\label{eq:L_sph}
\mathcal{L} = \sum_i \left\lbrace\frac{1}{2} I_i \left(\dot{\theta}_i^2 + \sin^2\theta_i \dot{\phi}^2\right) 
+ I_i \left(\dot{\psi}_i + \cos \theta_i \dot{\phi}_i \right)^2 \right\rbrace 
- U_\text{VR}(\cos\,\theta_{ij}(\theta_i-\theta_j,\phi_i-\phi_j))
\end{equation}
where
\begin{equation}\label{eq:thetaij}
	\cos\, \theta_{ij}  = \frac{1}{4}\left(\cos(\theta_i-\theta_j)(\cos(\phi_i-\phi_j)+1) - \cos(\theta_i+\theta_j)(\cos(\phi_i-\phi_j)-1) \right).
\end{equation}
We get the integrals of motion, $i=1,\ldots,N$,
\begin{equation}\label{eq:p_gen_s}
	s_i \equiv p_{\psi,i} = \frac{\partial \mathcal{L}}{\partial \dot{\psi}_i} =
	2 I_i (\dot{\psi}_i + \cos \theta_i \dot{\phi}_i) = 2I_i \omega_{\parallel,i} = \text{const.},
\end{equation}
which express the preservation of binding energies, that is of the planar kinetic energies $K_{\parallel,i}$. 
The generalized momenta for each body are 
\begin{equation}
\label{eq:p_gen}
	p_{\theta,i} \equiv \frac{\partial \mathcal{L}}{\partial \dot{\theta}_i} = I_i \dot{\theta}_i,
	\quad
	p_{\phi,i} \equiv \frac{\partial \mathcal{L}}{\partial \dot{\phi}_i} = \sin^2\theta_i I_i \dot{\phi}_i + s_i \cos\theta_i.  
\end{equation}

\subsection{Hamiltonian and Equations of Motion}\label{sec:Hamiltonian}

We are now able to write the general, canonical Hamiltonian of vector resonant relaxation
\begin{equation}\label{eq:H_VRR_sph}
	H(\{\theta,\phi,p_{\theta},p_{\phi}\}) = \sum_i \left\lbrace \frac{p_{\theta,i}^2}{2I_i}  + \frac{\left(p_{\phi,i} -s_i \cos\theta_i \right)^2}{2I_i \sin^2\theta_i} \right\rbrace  + U_\text{VR}(\{\theta,\phi\}).
\end{equation}
The potential energy $U_\text{VR}$ contains quadrapole terms and higher, as in (\ref{eq:U_VR}). The angle $\theta_{ij}$ between disks $i$ and $j$ is given in (\ref{eq:thetaij}).
The Hamilton equations of motion 
\begin{equation}
	\dot{p}_{\theta,i} = -\frac{\partial H_\text{VR}}{\partial \theta_i},\;	
	\dot{p}_{\phi,i} = -\frac{\partial H_\text{VR}}{\partial \phi_i}
\end{equation}
give the canonical system
\begin{align}
\label{eq:eom1}	&\dot{\theta}_i = \frac{1}{I_i}p_{\theta,i} ,\quad
	\dot{\phi}_i = \frac{1}{I_i\sin^2\theta_i} \left(p_{\phi,i} - s_i\cos\theta_i\right) \\
\label{eq:eom2}	&\dot{p}_{\theta,i} = \frac{\cos\theta_i\left(p_{\phi,i} - s_i\cos\theta_i\right)^2}{I_i\sin^3\theta_i} - \frac{s_i\left(p_{\phi,i} - s_i\cos\theta_i\right)}{I_i\sin\theta_i} - \frac{\partial U_\text{VR}}{\partial \theta_i}, \quad
	\dot{p}_{\phi,i} = - \frac{\partial U_\text{VR}}{\partial \phi_i}
\end{align}
and the equations of motion 
\begin{align}
	I_i\ddot{\theta}_i - I_i\sin\theta_i\cos\theta_i\dot{\phi}_i^2 + s_i\sin\theta_i\dot{\phi}_i  &= -\frac{\partial U_\text{VR}}{\partial \theta_i} \\
	I_i\sin^2 \theta_i\ddot{\phi}_i + 2I_i \sin\theta_i\cos\theta_i\dot{\theta}_i \dot{\phi}_i - s_i\sin\theta_i\dot{\theta}_i &= -\frac{\partial U_\text{VR}}{\partial \phi_i}.
	\end{align}
These may also be written in a useful vector form 
\begin{equation}\label{eq:eom_VRR}
	\dot{\bm{n}}_i = \bm{n}_i\times\left( \bm{\Omega}_i - \frac{I_i}{s_i}\ddot{\bm{n}}_i\right), 
\end{equation}
where
\begin{equation}
	\bm{\Omega}_i = \frac{1}{s_i}\sum_j \sum_l \mathcal{J}_{ij,l}\bm{n}_j  P_l'(\bm{n}_i\cdot \bm{n}_j).
\end{equation}
We used the chain rule $\partial/\partial\phi = (\partial \bm{n}/\partial\phi)(\partial (\bm{n}\cdot\bm{n}_j)/\partial \bm{n})(\partial/\partial (\bm{n}\cdot\bm{n}_j))$ and denote $P_l'(x) = dP_l(x)/dx$. For example, truncating at the quadrapole order we get
\begin{equation}\label{eq:Omega_quad}
	\bm{\Omega}_i = \frac{1}{s_i}\sum_j J_{ij} \bm{n}_j  (\bm{n}_i\cdot \bm{n}_j),
\end{equation}
where $J_{ij} = 3\mathcal{J}_{ij,2}$. 

\subsection{Phase Space Boundaries}\label{sec:Bounds}

In the following sections we will study the statistical mechanics generated by the canonical Hamiltonian (\ref{eq:H_VRR_sph}). We need therefore to determine the boundaries of phase space, compatible with the time-averaging procedure. Apparently, no bound needs to be imposed on the position angles $\theta\in [0,\pi]$, $\phi\in [0,2\pi]$. However, determining the value of bounds on generalized momenta requires careful investigation.

Firstly, we have to investigate the boundaries imposed by the dynamics itself. It shall be proven useful to rescale the physical parameters of the system. The strength of the interaction is determined by the two-body coupling constants $J_{ij}$. The physical properties of each disk are determined by two additional parameters; the moment of inertia $I_i$ and the spin magnitude $s_i$.
I introduce the arithmetic averages of these parameters 
\begin{equation}
		J = \frac{1}{N^2}\sum_{i> j} J_{ij} ,\quad
		I = \frac{1}{N}\sum_{i} I_{i}, \quad 
		s = \frac{1}{N}\sum_{i} s_{i},
\end{equation}		
which averages are used to scale each parameter as
\begin{equation}
	\tilde{J}_{ij} = \frac{J_{ij}}{J},\quad 
	\tilde{I}_i = \frac{I_i}{I},\quad
	\tilde{s}_i = \frac{s_i}{s}.
\end{equation}
The new `tilde' non-dimensional parameters are very useful in that for similar bodies, they all are of order 1. This property may allow us to distinguish between different families of bodies.

We inspect now the equations of motion (\ref{eq:eom_VRR}). Invoking the triangle inequality, we have at quadrapole order that 
\begin{equation}\label{eq:par_scale_O1}
	\left| \bm{\Omega}_i \right| = \left| \frac{1}{s_i}\sum_{j=1}^N J_{ij} \bm{n}_j  (\bm{n}_i\cdot \bm{n}_j)\right| \leq \frac{J N}{s}\frac{1}{\tilde{s}_i}\left(\frac{1}{N} \sum_{j=1}^N \tilde{J}_{ij}\right) \equiv t_c^{-1} \, \frac{\tilde{J}_i}{\tilde{s}_i},
\end{equation}
where $\tilde{J}_i = \sum_j \tilde{J}_{ij}/N$ is also of order one for a family of similar bodies. The timescale 
\begin{equation}\label{eq:t_c}
	t_\text{c} = \frac{s}{JN}.
\end{equation}
is induced naturally by the dynamics. Considering equations (\ref{eq:p_gen_s}), (\ref{eq:I_inertia_app}), (\ref{eq:om_par_app}) the spin scales as $s\sim m\sqrt{GMa}$. The gravitational coupling scales as $J\sim Gm^2/a$. Therefore $t_c$ scales like the apsidal precession timescale, that is the averaging timescale we use for VRR,
\begin{equation}\label{eq:t_c-av}
	t_c \sim  \frac{M}{mN}\sqrt{\frac{a^3}{GM}} \sim t_\text{aps} \equiv t_\text{av}.
\end{equation}
The general equations of motion (\ref{eq:eom_VRR}) may be rescaled in the completely equivalent form (with no additional assumption imposed) as 
\begin{equation}\label{eq:eom_tilde_gen}
	\frac{d\bm{n}_i}{d\tilde{t}} = \bm{n}_i\times\left( \tilde{\bm{\Omega}}_i - \varepsilon \,\frac{\tilde{I}_i}{\tilde{s}_i} \frac{d^2\bm{n}_i}{d\tilde{t}^2}\right),
\end{equation}
where time is scaled with the characteristic timescale 
\begin{equation}
	\tilde{t} = \frac{t}{t_\text{c}} .
\end{equation} 
We have denoted $\tilde{\bm{\Omega}}_i = t_c\bm{\Omega}_i$, giving at quadrapole order $\tilde{\bm{\Omega}}_i = \frac{1}{N}\sum_j \tilde{J}_{ij} \bm{n}_j  (\bm{n}_i\cdot \bm{n}_j)/\tilde{s}_i$. It is $|\tilde{\bm{\Omega}}_i| \leq \tilde{J}_i/\tilde{s}_i$.
In (\ref{eq:eom_tilde_gen}), naturally emerged the crucial, non-dimensional control parameter 
\begin{equation}\label{eq:epsilon}
	\varepsilon \equiv \frac{IJN}{s^2}.
\end{equation}
Since the moment of inertia (\ref{eq:I_inertia_app}) scales as $I\sim ma^2$, the control parameter scales as
\begin{equation}
	\varepsilon \sim \frac{mN}{M},
\end{equation}
where $M = M_\star + M_\bullet$. Therefore $\varepsilon^{-1}$ measures the intensity of gravitation of the central mass $M_\bullet$ with respect to that of the cluster $M_\star$. It is $\varepsilon < 1$.

The general equations of motion (\ref{eq:eom_tilde_gen}) impose consequently (at least at quadrapole order) the following inequality for all disks $i$
\begin{equation}\label{eq:omega_bound}
	\left|\frac{d\bm{n}_i(t)}{d \tilde{t}}\right| 
	\leq  \frac{\tilde{J}_i}{\tilde{s}_i} +
	\varepsilon A_i(t) , \quad \forall i=1,\ldots,N.
\end{equation}
where we denote $A_i$ the non-dimensional angular acceleration 
\begin{equation}
	A_i(t) \equiv   \frac{\tilde{I}_i}{\tilde{s}_i} \left|\frac{d^2\bm{n}_i(t)}{d\tilde{t}^2}\right|.
\end{equation}

All calculations up to here have been performed on the general equations of motion (\ref{eq:eom_VRR}) without imposing any condition. The constraint (\ref{eq:omega_bound}) is naturally satisfied by the dynamics itself. Now, I wish to inspect to what extent is the time-averaging procedure of VRR combatible with the induced dynamics. The identification (\ref{eq:t_c-av}) of the characteristic timescale with the averaging timescale allows us to write the extreme condition of validity of the averaging procedure
\begin{equation}\label{eq:omega_bound_VRR}
	\left|\frac{d\bm{n}_i(t)}{d\tilde{t}}\right|_\text{VRR} \leq 1,
\end{equation}
for all orbits. This expresses the fact that the angular velocity of the orbital planes $|\bm{\omega}_\perp |= |d\bm{n}/dt|$ cannot be higher than the frequency at which the planes themselves are defined. Orelse the definition of such an orbital plane fails.

According to the implicit bound (\ref{eq:omega_bound_VRR}), the VRR-constraint (\ref{eq:omega_bound_VRR}) is satisfied for a system subject to an evolution which satisfies at all $t > t_c$ that 
\begin{equation}\label{eq:A_con}
	\frac{\tilde{J}_i}{\tilde{s}_i} +
	\varepsilon A_i(t) \lesssim 1.
\end{equation}
For a family of similar orbits, this is satisfied if $A_i(t) < \varepsilon^{-1}$ for all $t>t_c$.

For $\varepsilon \ll 1$ this translates simply to non-diverging accelerations and therefore (\ref{eq:omega_bound_VRR}) is then typically satisfied. Note that the general evolution (\ref{eq:eom_tilde_gen}) becomes at zeroth order of $\varepsilon$.
\begin{equation}\label{eq:eom_VRR_SO3}
	\frac{d\bm{n}_i}{dt} = \bm{n}_i\times\bm{\Omega}_i, \quad \varepsilon \rightarrow 0. 
\end{equation}
Such type of equations of motion for VRR were firstly suggested in \cite{2015MNRAS.448.3265K}, where they were generated directly from the potential energy term applying the $SO(3)$ Poisson algebra satisfied by the components of $\bm{n}_i$ without any reference to a kinetic term. 

In addition, it might very well be the case that $A_i(t) < \varepsilon^{-1}$, for $\varepsilon$ of order one or smaller, at the timescales involved and for astrophysical systems of interest (this depends on the physical conditions at timescales $t < t_c$). Then, the constraint (\ref{eq:omega_bound_VRR}) is satisfied. Such a case requires further investigation, but it is possible, and is supported by numerical simulations \cite{Meiron_2018arXiv180607894M} of Globular Clusters. Certainly, the lower that $\varepsilon$ is, the more consistently in time the VRR-constraint (\ref{eq:A_con}) is expected to be satisfied and for more orbits. 
I will consider in our statistical mechanics' analysis all values
\begin{equation}
	\varepsilon < 1.
\end{equation}
If $\varepsilon \ll 1$ the central massive object $M_\bullet$ dominates completely the gravitational potential. As $\varepsilon$ gets closer to one, the mass of the cluster $M_\star$ becomes more comparable to the mass of the central massive object and for $\varepsilon \sim 1$ the later is totally absent. 
I stress once more that the analysis is valid, for $\varepsilon$ not negligible at zeroth order, only for these initial conditions $t<t_c$, that for the vast majority of orbits the constraint (\ref{eq:A_con}) is satisfied at $t>t_c$ and at least up to times $t\sim t_\text{VRR}$, when thermal equilibrium is expected to be achieved. Even in the case $\varepsilon \ll 1$ we will be able to quantify the effect of different $\varepsilon$ values no matter how small the later is assumed to be.

We have to translate the constraint (\ref{eq:omega_bound_VRR}) to boundaries of our  canonical phase space. The normal angular velocity of the disks $\bm{\omega}_\perp = d\bm{n}/dt$ is related with the generalized momenta (\ref{eq:p_gen}) as
\begin{equation}\label{eq:dn_p_gen}
	\left|\frac{d\bm{n}_i}{dt}\right|^2 \equiv \dot{\theta}_i^2 + \sin^2\theta_i\dot{\phi}_i^2
	=  \frac{1}{\varepsilon^2 t_c^2} \left(\frac{\tilde{p}_{\theta,i}^2}{\tilde{I}_i^2} +  
	\frac{(\tilde{p}_\phi - \tilde{s}_i \cos\theta_i)^2}{\tilde{I}_i^2 \sin\theta_i^2}\right),
\end{equation}
where
\begin{equation}
	\tilde{p}_{\theta,i} = \frac{p_{\theta,i}}{s},\quad
	\tilde{p}_{\phi,i} = \frac{p_{\phi,i}}{s}.	
\end{equation}

Combining (\ref{eq:dn_p_gen}) with (\ref{eq:omega_bound_VRR}) we get
\begin{equation}\label{eq:p_constraint}
\left(\frac{\tilde{p}_{\theta,i}}{\tilde{I}_i}\right)^2 + \left( \frac{\tilde{p}_{\phi,i} - \tilde{s}_i \cos\theta_i}{\tilde{I}_i\sin\theta_i}\right)^2  \leq \varepsilon^2.
\end{equation}
This bound (\ref{eq:p_constraint}) defines the region of the canonical phase space, that is accessible by the generalized momenta.
The accessible values of the $i$th generalized momenta form an elliptical disk $D$ in phase space. 

Besides the control parameter $\varepsilon$, another key quantity for our analysis is the angular velocity $\bm{\omega}_\perp = d\bm{n}/dt$, as in (\ref{eq:omega_def}). It describes the magnitude of both precession and nutation, that is the magnitude of angular velocity of the effective disks (orbital planes) rotating about any of their diameters (excluding spinning). Its dispersion may be expected to attain a constant value at thermal equilibrium at times $t\sim t_\text{VRR}\gg t_c$. 
We rescale $\bm{\omega}_\perp$ by $t_c$ and introduce the quantity 
\begin{equation}\label{eq:omega_resc}
	\omega_i \equiv |t_c\bm{\omega}_{\perp,i}| = \left|\frac{d{\bm{n}_i}}{d\tilde{t}}\right|
	= \frac{1}{\varepsilon}\left\lbrace \left( \frac{\tilde{p}_{\theta,i}}{\tilde{I}_i}\right)^2 + \left( \frac{\tilde{p}_{\phi,i} - \tilde{s}_i \cos\theta_i}{\tilde{I}_i\sin\theta_i}\right) \right\rbrace^{1/2},
	\quad 0 \leq \omega_i \leq 1.
\end{equation}
Next, I introduce the transformations
\begin{align}\label{eq:tra_p}
\begin{split}
	\tilde{p}_{\theta,i}(\omega_i,u_i,\theta'_i) &= \tilde{I}_i\varepsilon \omega_i\cos u_i, \\
	\tilde{p}_{\phi,i}(\omega_i,u_i,\theta'_i) &= \tilde{I}_i\varepsilon \omega_i\sin\theta'_i \sin u_i + \tilde{s}_i\cos\theta'_i, \\
	\theta_i(\omega_i,u_i,\theta'_i) &= \theta'_i,
\end{split}
\end{align}
with $u_i\in [0,2\pi]$. 
The $\theta'$ is introduced to allow us consistently calculate the Jacobian of the transformation, which equals $\varepsilon^2 \tilde{I}_i^2\omega_i$. 
I stress that the canonical phase space element is
\begin{equation}
	d\tau = \prod_{i=1}^N d\theta_i \,d\phi_i \, d\tilde{p}_{\theta,i}\, d\tilde{p}_{\phi,i},
\end{equation}
which does not include the solid angle. The later will naturally emerge by the transformation (\ref{eq:tra_p}).
Indeed, the total phase space volume element, including the limits of any integration, is transformed under (\ref{eq:tra_p}) as
\begin{equation}\label{eq:dtau}
	\int_0^{2\pi}\int_0^\pi  \iint_D  \prod_{i=1}^N d\phi_i d\theta_{i}\,d\tilde{p}_{\theta,i} d\tilde{p}_{\phi,i} \,\{ \cdots\}
	= \int_0^{2\pi}\int_0^\pi \int_0^{2\pi}  \int_0^1 \prod_{i=1}^N \varepsilon^2\tilde{I}_i^2\omega_i d\mathcal{A}_i du_i d\omega_i \, \{ \cdots\},
\end{equation}
where I mean an $N$-integral for each integration symbol.
I denote $d\mathcal{A}_i$ the solid angle
\begin{equation}
	d\mathcal{A}_i \equiv \sin\theta_i d\theta_i d\phi_i.
\end{equation}

Note that according to the definition of generalized momenta (\ref{eq:p_gen}), the transformations (\ref{eq:tra_p}) give for the nutation and precession of orbital planes, respectively,
\begin{align}
\label{eq:nut}
	\frac{d \theta_i}{d\tilde{t}} &= \omega_i (\tilde{t}) \cos u_i(\tilde{t}), \\ 
\label{eq:prec}
	\frac{d \phi_i}{d\tilde{t}} &= \omega_i (\tilde{t}) \frac{ \sin u_i(\tilde{t})}{\sin\theta_i(\tilde{t})}.
\end{align}
These expressions will allow us to calculate ensemble averages of $\dot{\theta}$, $\dot{\phi}$ over the transformed phase space element (\ref{eq:dtau}).

For reasons of completeness, I will consider also  the case of continuous rigid bodies in section \ref{sec:RB}. These are not subject to (\ref{eq:omega_bound_VRR}) and therefore allowed to probe the whole phase space. The $\varepsilon_{\text{RB}}$ of continuous rigid bodies is finite and given by equation (\ref{eq:epsilon}).

\section{Equilibrium Statistical Mechanics of Vector Resonant Relaxation}\label{sec:Stat}

\subsection{Microcanonical Ensemble}

In the microcanonical ensemble I impose conservation of the time-averaged energy (\ref{eq:H_VRR_sph})
\begin{equation}\label{eq:E_constraint}
	H\left(\left\lbrace \theta,\phi, p_\theta, p_\phi \right\rbrace \right) = E.
\end{equation}
It expresses the exchange between the effective disks of gravitational dynamical energy and rotational kinetic energy about any diameter of a disk. Therefore, it expresses the relaxation to thermal equilibrium of their angular velocities along with their orientation.

In addition to the energy constraint (\ref{eq:E_constraint}), the system is subject to the conservation of total spin (related to the sum of planar angular momentum of each mass $m_i$)
\begin{equation}\label{eq:L_constraint}
	\bm{S} = \sum_i^N s_i \bm{n}_i = const. 
\end{equation}
For simplicity, from now on I denote 
\begin{equation}
	K \equiv  \sum_i \left\lbrace \frac{p_{\theta,i}^2}{2I_i}  + \frac{\left(p_{\phi,i} -s_i \cos\theta_i \right)^2}{2I_i \sin^2\theta_i} \right\rbrace ,
	\quad
	U \equiv U_\text{VR}.
\end{equation}

I define a \textit{microcanonical ensemble}, as the ensemble of microstates corresponding to a certain macrostate $(E,\bm{S})$. I use the 4D phase-space volume 
\begin{equation}\label{eq:Omega_PS}
	\Omega (E,\bm{S}) = \frac{1}{b^{2N}N!}\int d^{N}\theta\, d^{N}\phi\, d^{N}p_\theta\, d^Np_\phi\, \delta(E - H(\{ \theta,\phi,p_\theta,p_\phi\}) \delta ( \bm{S} - \sum_i s_i \bm{n}_i(\theta_i,\phi_i)).
\end{equation}
I define the entropy of the system as (not to be confused with spin vector $\bm{S}$ and its magnitude)
\begin{equation}\label{eq:S}
	\mathcal{S}_B (E,\bm{S}) = k\ln \Omega (E,\bm{S}).
\end{equation}
I emphasize that the correct phase space volume element is $d^{N}\theta\, d^{N}\phi\, d^{N}p_\theta\, d^Np_\phi$ induced by the relevant canonical variables. 

\subsection{Canonical Ensemble}\label{sec:Stat_Can}
Next, I introduce the following Laplace transform of $\Omega (E,\bm{S})$
\begin{equation}\label{eq:Z_laplace}
	Z(\beta, \bm{S}) \equiv \int dE\, \Omega(E,\bm{S}) e^{-\beta E}
\end{equation}
which allows us to define a \textit{canonical ensemble} as the ensemble of microstates corresponding to certain values of $(\beta, \bm{S})$. The macroscopic variable $\beta$ is the conjugate variable to $E$ in the sense of (\ref{eq:Z_laplace}). The Helmholtz free energy is defined as
\begin{equation}\label{eq:F_def}
	F (\beta,\bm{S}) = -\frac{1}{\beta} \ln Z (\beta,\bm{S}).
\end{equation}

The kinetic energy may be written in the rescaled parameters which I introduced in section \ref{sec:Bounds}
\begin{equation}\label{eq:K_resc}
	K \equiv  JN \sum_i \frac{1}{2\varepsilon } \left( \frac{\tilde{p}_{\theta,i}^2}{\tilde{I}_i}  + \frac{\left(\tilde{p}_{\phi,i} - \tilde{s}_i \cos\theta_i \right)^2}{ \tilde{I}_i\sin^2\theta_i} \right) = JN \sum_i \frac{\varepsilon}{2}\tilde{I}_i\omega_i^2 .
\end{equation}
Introducing in addition the non-dimensional quantity
\begin{equation}\label{eq:b_tilde}
	\tilde{\beta } = JN\beta,
\end{equation}
the canonical partition function (\ref{eq:Z_laplace}) may be written 
\begin{align}
	Z &= \frac{1}{N!}\int_0^\pi\int_0^{2\pi}\iint_D \prod_{i=1}^N d\theta_i\, d\phi_i\, 
	\left\lbrace d\tilde{p}_{\theta,i}\, d\tilde{p}_{\phi,i} \,  
	  e^{- \tilde{\beta}\sum \frac{1}{2\varepsilon } \left( \frac{\tilde{p}_{\theta,i}^2}{\tilde{I}_i}  + \frac{\left(\tilde{p}_{\phi,i} - \tilde{s}_i \cos\theta_i \right)^2}{ \tilde{I}_i\sin^2\theta_i} \right) }
	  \right\rbrace 
	e^{- \beta \,U } \delta (\bm{S} - \sum_i s_i \bm{n}_i),\\
	\label{eq:Z_can_om}
	&= \frac{1}{N!}
	\int_0^\pi \int_0^{2\pi}
	\prod_{i=1}^N d\mathcal{A}_i\,
	\left\lbrace \int_0^{2\pi} du_i \int_0^
	1 d\omega_i \,\varepsilon^2 \tilde{I}_i^2\,\omega_i\,
	  e^{-\sum \frac{ \tilde{\beta} \varepsilon}{2} \tilde{I}_i\omega_i^2} 
	  \right\rbrace 
	e^{- \beta\,U }\delta (\bm{S} - \sum_i s_i \bm{n}_i),
\end{align}
where I invoke the transformations (\ref{eq:tra_p}), which suggest the use of phase space volume (\ref{eq:dtau}).
The integration over each $d\omega_i$, $du_i$ can be straightforwardly preformed
\begin{equation}
	\int_0^{2\pi} du_i \int_0^1 d\omega_i \,\varepsilon^2 \tilde{I}_i^2\,\omega_i\,
	  e^{-\frac{\tilde{\beta} \varepsilon}{2} \tilde{I}_i\omega_i^2 }
	  =
\frac{2\pi \varepsilon}{\tilde{\beta}} 
\tilde{I}_i \left( 1 - e^{-\frac{\tilde{\beta}\varepsilon}{2} \tilde{I}_i}\right).
\end{equation}

Finally, the canonical partition function is written as
\begin{equation}\label{eq:Z_can}
	Z = \frac{1}{N!}
	\left(\frac{2\pi\varepsilon}{\tilde{\beta}}\right)^N
	\int \prod_{i=1}^N d\mathcal{A}_i\, h_i(\varepsilon\tilde{\beta}) \,  
	e^{- \beta \,U} \delta (\bm{S} - \sum_i s_i \bm{n}_i ),
\end{equation}
where
\begin{equation}\label{eq:h_i}
	h_i(\varepsilon\tilde{\beta}) =
	\tilde{I}_i \left( 1 - e^{-\frac{\tilde{\beta}\varepsilon}{2} \tilde{I}_i}\right).
\end{equation}

\subsection{Gibbs-canonical Ensemble}

Finally, I define the \textit{Gibbs-canonical ensemble}. I introduce the following Laplace transform of $Z (\beta,\bm{S})$
\begin{equation}\label{eq:Xi_laplace}
	\Xi  (\beta,\bm{\gamma}) \equiv \int d\bm{S}\, Z (\beta,\bm{S}) e^{- \bm{\gamma}\cdot \bm{S}}
	 = \iint d\bm{S}\, dE\, \Omega(E,\bm{S}) e^{-\left(\beta E + \bm{\gamma} \cdot \bm{S} \right)}.
\end{equation}
The parameter $\bm{\gamma}$ is the variable conjugate to the spin $\bm{S}$ and acquires dimensions of angular velocity over energy. One may associate with $\bm{\gamma} \cdot d\bm{S}/\beta $ a quantity of work performed by the system because of external disturbances, similarly to the work $PdV$ done under constant pressure for gases.

Substituting Eq. (\ref{eq:Z_can}) into (\ref{eq:Xi_laplace}), we get
\begin{equation}\label{eq:Xi}
	\Xi (\beta,\bm{\gamma}) = \frac{1}{N!}\left(\frac{2\pi\varepsilon}{\tilde{\beta} } \right)^N\, 
	\int\prod_i d\mathcal{A}_i h_i(\beta) \,e^{- \left( \beta\, U +  \bm{\gamma}\cdot \bm{S}\right)},
\end{equation}
where $\bm{S}$ satisfies here
\begin{equation}
	\bm{S} = \sum_{i=1}^N s_i\bm{n}_i.
\end{equation}

I define the Gibbs free energy
\begin{equation}\label{eq:bG_can}
	G (\beta,\bm{\gamma}) = - \frac{1}{\beta}\ln \Xi (\beta,\bm{\gamma}).
\end{equation}
We get
\begin{equation}\label{eq:G_free}
	\beta G (\beta,\bm{\gamma}) = \frac{N}{\tilde{\beta}} + \xi (\beta, \bm{\gamma}) -\ln\frac{(2\pi \varepsilon)N}{N!},
\end{equation}
where 
\begin{equation}
	\xi (\beta, \bm{\gamma})= - \ln\int\prod_i d\mathcal{A}_i h_i(\beta) \,e^{- \left( \beta\, U +  \gamma \cdot \bm{S} \right)}
\end{equation}

Definitions (\ref{eq:Xi_laplace}) and (\ref{eq:bG_can}) suggest that the spin as an ensemble mean may be given by
\begin{equation}\label{eq:L_Gcan_mean}
	\LA \bm{S}\RA_\text{ens} = \frac{\partial(\beta G)}{\partial \bm{\gamma}},
\end{equation}
where it is 
\begin{equation}\label{eq:L_mean_Gibbs}
	\LA \bm{S}\RA_\text{ens} \equiv \frac{ \int d\bm{S}\, \bm{S}\, \int dE\, \Omega(E,\bm{S}) e^{-\left(\beta E + \bm{\gamma} \cdot \bm{S}\right)} }{ \int d\bm{S}\, \int dE\, \Omega(E,\bm{S}) e^{-\left( \beta E + \bm{\gamma} \cdot \bm{S}\right)} } .
\end{equation}
Incorporating eqs (\ref{eq:G_free}) we get 
\begin{equation}\label{eq:S_mean_fU}
	\LA \bm{S}\RA = \int \prod_i d\mathcal{A}_i \,\bm{S} \,f_{A}(\theta_i,\phi_i),
\end{equation}
where the spacial, $N$-particle angular distribution function is 
\begin{equation}\label{eq:f_A}
	f_{A}(\theta_i,\phi_i) = \frac{h_i(\beta) \,e^{- \left( \beta\, U  +  \bm{\gamma}\cdot \bm{S} \right)}}{\int\prod_j d\mathcal{A}_j h_j(\beta) \,e^{- \left( \beta\, U  +  \bm{\gamma}\cdot \bm{S} \right)} }.
\end{equation}

Definitions (\ref{eq:Xi_laplace}) and (\ref{eq:bG_can}) also suggest that the energy, as an ensemble mean, can be calculated as
\begin{equation}
	\LA E \RA_\text{ens} = \frac{\partial (\beta G)}{\partial \beta}.
\end{equation}
Applying again (\ref{eq:G_free}) we get
\begin{equation}\label{eq:energy_mean_ens}
	\LA E \RA_\text{ens} = \LA K\RA_\text{ens} + \LA U \RA_\text{ens} ,
\end{equation}
where the potential energy is
\begin{equation}\label{eq:U_ens}
	\LA U\RA_\text{ens} = \int \prod_i d\mathcal{A}_i\, U  \,f_{A}(\theta_i,\phi_i),
\end{equation}
and the kinetic energy is
\begin{equation}\label{eq:K_VRR}
	\LA K\RA_\text{ens} = NkT\left\lbrace 1 - \tilde{\beta} \frac{1}{N}\sum_{i=1}^N\frac{\partial \ln h_i}{\partial \tilde{\beta}} \right\rbrace.
\end{equation}
The typical kinetic energy term $NkT$, reflects the presence of two kinetic degrees of freedom, namely, precession and nutation of orbital planes. The corrections to this term, due to the VRR phase space bound (\ref{eq:p_constraint}), are governed by $\varepsilon\tilde{\beta}$ (not $\varepsilon$ itself) and also by the moment of inertia $\tilde{I}_i=I_i/I$, because of the function $h_i(\varepsilon\tilde{\beta};\tilde{I}_i)$ as in Eq. (\ref{eq:h_i}). 
 I will inspect in detail the VRR kinetic energy (\ref{eq:K_VRR}) in the separate section (\ref{sec:K_VRR}).

Let us first define a thermodynamic limit.
The Gibbs partition function 
\begin{equation}
	\Xi  (\beta,\bm{\gamma} ) = \iint dE d\bm{S}\, \Omega(E,\bm{S}) e^{-\beta (E + \bm{\gamma}\cdot \bm{S})}
	= \iint dE d\bm{S}\, e^{-N(-\frac{\mathcal{S}_B(E,\bm{S})}{kN} + N\beta\frac{E}{N^2} +  \bm{\gamma} \frac{\bm{S}}{N})},
\end{equation}
may be calculated by use of the saddle point method for $N\rightarrow \infty$, 
provided that the following dimensionless variables (be careful not to confuse the entropy $\mathcal{S}_B$ with spin) are held constant 
\begin{equation}\label{eq:therm_limit}
	\tilde{\mathcal{S}}_B = \frac{\mathcal{S}_B}{k N} ,\quad
	\tilde{\beta} = J N\beta ,\quad
	\tilde{E} = \frac{E}{JN^2} ,\quad
	\tilde{\bm{\gamma}} = s\bm{\gamma},\quad
 \tilde{\bm{S}} = \frac{\bm{S}}{N s}.
\end{equation}
Note that $JN\beta$ and $E/JN^2$ are the analogues of the variables $(Gm^2/R) N\beta$ and $E/(Gm^2/R)N^2$ used to define a proper thermodynamic limit for the $2$-body relaxation of the self-gravitating gas \cite{2002NuPhB.625..409D}.
We may write
\begin{equation}\label{eq:Xi_tilde}
	\Xi  (\beta,\bm{\gamma}) = s J N^3 \iint d\tilde{E}d\tilde{\bm{S}} \, e^{-N\left(-\tilde{\mathcal{S}}_B + \tilde{\beta} \tilde{E} + \tilde{\bm{\gamma}}\cdot \tilde{\bm{S}}\right) }
\end{equation}
and now provided that $\tilde{\mathcal{S}}_B, \tilde{\beta}, \tilde{E}, \tilde{\bm{\gamma}}, \tilde{\bm{S}}$ are held constant in the limit $N\rightarrow \infty$ the integral is dominated by the minimum of the exponent.
Therefore, by Eq. (\ref{eq:bG_can}) $\Xi = \exp(-\beta G)$, we have up to additive constants
\begin{equation}\label{eq:G_tilde_1}
	\tilde{\beta} \tilde{G}(\tilde{\beta},\tilde{\bm{\gamma}}) = \inf_{\tilde{E},\tilde{\bm{S}}} (-\tilde{\mathcal{S}}_B(\tilde{E},\tilde{\bm{S}}) + \tilde{\beta} \tilde{E} + \tilde{\bm{\gamma}} \tilde{\bm{S}}),
\end{equation}
where
\begin{equation}
	\tilde{G}(\tilde{\beta},\tilde{\bm{\gamma}}) = \frac{G( \beta,\tilde{\bm{\gamma}}) }{JN^2}.
\end{equation}
The values $\tilde{E}_\text{e}$, $\tilde{\bm{S}}_\text{e}$ satisfying the minimum of (\ref{eq:G_tilde_1}) define the stable statistical equilibrium of the system for fixed values of $\tilde{\beta}$, $\tilde{\bm{\gamma}}$. For these values we can write
\begin{equation}\label{eq:G_Legendre_1}
	\tilde{\beta} \tilde{G}(\tilde{\beta},\tilde{\bm{\gamma}} ) = -\tilde{\mathcal{S}}_B(\tilde{E}_\text{e}(\tilde{\beta},\tilde{\bm{\gamma}}),\tilde{\bm{S}}_\text{e}(\tilde{\beta},\tilde{\bm{\gamma}})) + \tilde{\beta} \tilde{E}_\text{e}(\tilde{\beta},\tilde{\bm{\gamma}}) + \tilde{\bm{\gamma}} \tilde{\bm{S}}_\text{e}(\tilde{\beta},\tilde{\bm{\gamma}}).
\end{equation}

Similarly, working with the canonical partition function (\ref{eq:Z_laplace}) and defining
\begin{equation}
	\tilde{F}(\tilde{\beta},\tilde{\bm{S}}) = \frac{F(\beta,\bm{S})}{JN^2}.
\end{equation}
we get
\begin{equation}\label{eq:F_tilde}
	\tilde{\beta} \tilde{F}(\tilde{\beta},\tilde{\bm{S}}) = \inf_{\tilde{E}} (-\tilde{\mathcal{S}}_B(\tilde{E},\tilde{\bm{S}}) + \tilde{\beta} \tilde{E} ).
\end{equation}
This reveals that $\tilde{\beta}\tilde{F}(\tilde{\beta},\tilde{\bm{S}})$ is the Legendre-Fenchel transform \cite{Campa_2014} of $\tilde{\mathcal{S}}_B(\tilde{E},\tilde{\bm{S}})$ with respect to the energy.
The value $\tilde{E}_\text{e}$ satisfying the minimum of (\ref{eq:F_tilde}) defines the stable statistical equilibrium of the system for fixed values of $\tilde{\beta}$ and $\tilde{\bm{S}}$ (our canonical ensemble). For this value we can write
\begin{equation}\label{eq:F_Legendre}
	\tilde{\beta} \tilde{F}(\tilde{\beta},\tilde{\bm{S}}) = -\tilde{\mathcal{S}}_B(\tilde{E}_\text{e}(\tilde{\beta},\tilde{\bm{S}}),\tilde{\bm{S}}) + \tilde{\beta} \tilde{E}_\text{e}(\tilde{\beta},\tilde{\bm{S}}).
\end{equation}
The equilibria defined by either minima of free energy (\ref{eq:F_tilde}) or maxima of entropy are identical. Nevertheless, for long-range interacting systems, the stability properties of these equilibria under conditions of constant energy (microcanonical formulation) are different than their stability under conditions of constant temperature (canonical formulation) during a phase transition. Mathematically, this is because of the irreverisbility of the Legendre-Fenchel transform (\ref{eq:F_Legendre}) in this case. Physically, this is because it is impossible for the system to achieve phase separation during the phase transition due to the non-additivity of energy. The phase transition proceeds in the canonical ensemble by jumping from one phase to the other through out-of-equilibrium states, while this region is replaced by stable states with negative heat capacity in the microcanonical case (physically, constant energy conditions). This is further discussed in Section \ref{sec:ineq_ens}, following \cite{Campa_2014}.

Finally, let us provide the transform between canonical and Gibbs-canonical ensembles. We may write Eq. (\ref{eq:G_tilde_1}) as
\begin{equation}
	\tilde{\beta} \tilde{G}(\tilde{\beta},\tilde{\bm{\gamma}}) = \inf_{\tilde{\bm{S}}} \left( \inf_{\tilde{E}}( -\tilde{\mathcal{S}}_B(\tilde{E},\tilde{\bm{S}}) + \tilde{\beta} \tilde{E}) + \tilde{\bm{\gamma}} \tilde{\bm{S}}\right) ,
\end{equation}
which by use of (\ref{eq:F_tilde}) becomes
\begin{equation}\label{eq:G_tilde}
	\tilde{G}(\tilde{\beta},\tilde{\bm{\gamma}}) = \inf_{\tilde{\bm{S}}} ( \tilde{F}(\tilde{\beta},\tilde{\bm{S}}) + \tilde{\bm{\gamma}}\tilde{\bm{S}} ) .
\end{equation}
which reveals that $\tilde{G}(\tilde{\beta},\tilde{\bm{\gamma}})$ is the Legendre-Fenchel transform of $\tilde{F}(\tilde{\beta},\tilde{\bm{S}})$ with respect to the spin.
By use of Eqs. (\ref{eq:G_Legendre_1}), (\ref{eq:F_Legendre}) we get
\begin{equation}\label{eq:G_Legendre}
	\tilde{G}(\tilde{\beta},\tilde{\bm{\gamma}}) = \tilde{F}(\tilde{\beta},\tilde{\bm{S}}_\text{e}(\tilde{\beta},\tilde{\bm{\gamma}})) + \tilde{\bm{\gamma}}\tilde{\bm{S}}_\text{e}(\tilde{\beta},\tilde{\bm{\gamma}}).
\end{equation}
Because of the additivity of spin it is not expected to appear inequivalence between canonical and Gibbs-canonical ensembles. However, there is inequivalence between Gibbs-canonical and microcanonical ensembles of the similar nature as that between canonical and microcanonical cases.

\section{Inequivalence of Ensembles}\label{sec:ineq_ens}

According to Eqs. (\ref{eq:F_tilde}), (\ref{eq:F_Legendre}), the function $\tilde{\beta}\tilde{F}$ is the Legendre-Fenchel transform of normalized entropy $\tilde{\mathcal{S}}_B$ with respect to energy.
In order for microcanonical and canonical ensembles to be equivalent in the thermodynamic limit, this transform should be invertible for all $\tilde{E}$, i.e. the function $\tilde{E} = \tilde{E}(\tilde{\beta},\tilde{\bm{S}})$ should be invertible with respect to $\tilde{\beta}$.
However, this is not guaranteed if the normalized entropy is not a strictly concave function as emphasized in Ref. \cite{Campa_2014}. The reason these authors indicate is that the Legendre-Fenchel transform is by definition a concave function. Therefore the inverse transform of (\ref{eq:F_tilde})
\begin{equation}
\tilde{\mathcal{S}}_B'(\tilde{E},\tilde{\bm{S}})  = \inf_{\tilde{\beta}} (-\tilde{\beta}\tilde{F}(\tilde{\beta},\tilde{\bm{S}}) + \tilde{\beta} \tilde{E} )
\end{equation}
is concave by definition. If for some range of parameters $\tilde{S}$ is not concave, then for sure
\begin{equation}
\tilde{\mathcal{S}}_B' \neq \tilde{\mathcal{S}}_B,\quad\text{when}\; \tilde{\mathcal{S}}_B\;\text{not concave},
\end{equation}
and the ensembles are not equivalent. This occurs in long-range interacting systems, like the one I investigate here, undergoing a phase transition. Then, there appears a convex `intruder' in the specific entropy $\tilde{\mathcal{S}}_B$ which in the case of short-range systems is replaced by its concave envelope \cite{Campa_2014}, restoring the ability to perform an inverse Legendre transform. This envelope corresponds to a phase separation. However, this operation cannot be performed if the energy of the system is \textit{non-additive}, that is the case of long-range interacting systems. Therefore, during a phase transition the equivalence of ensembles breaks down in such systems. States that are not stable in the canonical ensemble, i.e. under physical conditions of constant temperature, are stable in the microcanonical ensemble, i.e. under conditions of constant energy. The microcanonical specific heat \cite{Campa_2014}
\begin{equation}
C_\text{mic.} = \frac{\partial E}{\partial T} = -\frac{1}{T^2}\frac{\partial^2 S}{\partial E^2} < 0,\quad
\text{if}\; \frac{\partial^2 \mathcal{S}_B}{\partial E^2} >0,
\end{equation}
is negative in the convex region $\partial^2 S/\partial E^2 >0$ (recall that due to the non-additivity it is impossible to define a concave envelope of higher entropy). However, the canonical specific heat is always positive
\begin{equation}
C_\text{can.} = \frac{\partial \LA E\RA_\text{ens} }{\partial T} = \frac{1}{T^2} \LA (\Delta E)^2\RA_\text{ens} > 0.
\end{equation}

Equivalently, one may consider that the additional constraint $E=\text{const.}$ of the microcanonical ensemble prevents the modes which destabilize the system in the canonical ensemble (presence of a heat bath) from developing. In a short range system these variations would guide the system towards the state of phase separation in both ensembles, which state nevertheless does not exist in a long-range interacting system, because of energy's non-additivity. The long-range interacting system will stay trapped in a negative specific-heat state when energy is preserved in a phase-transition region.

\section{Gravitational Phase Transitions for Equal Couplings}\label{sec:GPT}

Here I briefly review the VRR gravitational phase transitions, discovered firstly for zero kinetic term in \cite{2017ApJ...842...90R}. It is instructive for the rest of the paper to identify the possible spacial equilibrium distributions. More importantly, I will show that the kinetic term has not any effect in the spacial equilibrium distributions of any family of disks. Therefore the VRR gravitational phase transitions do hold and remain unaltered for any value of $\varepsilon \sim mN/M <1$.

I will assume equal couplings and spins
\begin{equation}\label{eq:param_MF}
	\tilde{J}_{ij} = 1,\quad \tilde{s}_i = 1.
	\quad \forall i,j=1,\ldots, N.
\end{equation}
For now I assume for simplicity also $\tilde{I}_i=1$, but the moments of inertia may not be equal and the analysis of this section still holds, as we will discuss later in this section.
The condition (\ref{eq:param_MF}) accounts for a single family of orbits with very similar properties.
I also truncate the interaction energy $U_\text{VR}$ (\ref{eq:U_VR}) at the quadrapole denoting it
\begin{equation}
	U = -\frac{J}{2} \sum_{i\neq j} \left((\bm{n}_i\cdot\bm{n}_j)^2 - \frac{1}{3}\right).
\end{equation}
I shall calculate the distribution function and the self-consistency equations for the order parameters, which need to be identified. 

\begin{figure}[tbp]
\begin{center}
        \hspace{-0.7cm}\shortstack{\includegraphics[scale = 0.25]{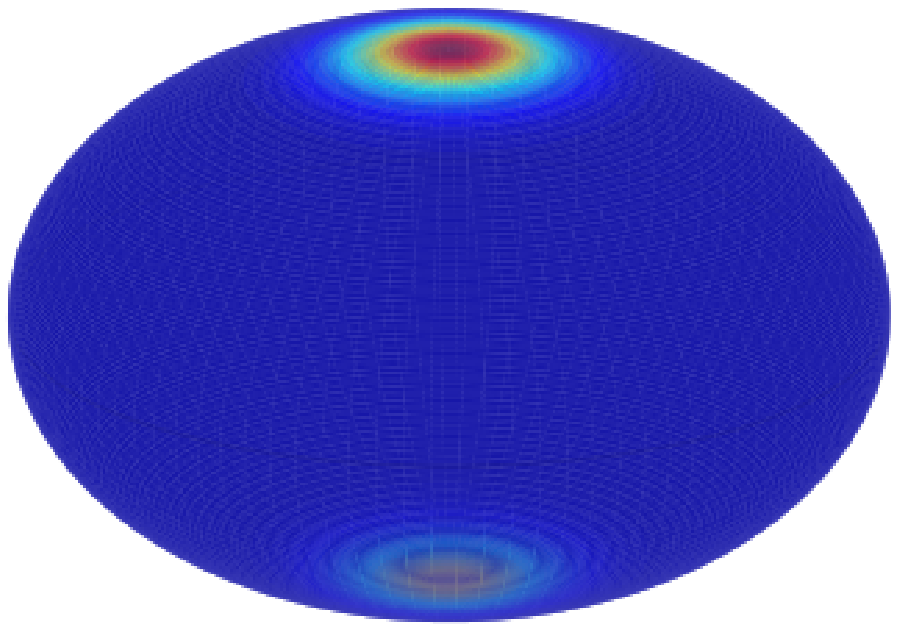}\\[-4ex]\hspace{0.7em}\normalsize{$\Sigma_1$}}
        \hspace{-0.7cm}\shortstack{\includegraphics[scale = 0.25]{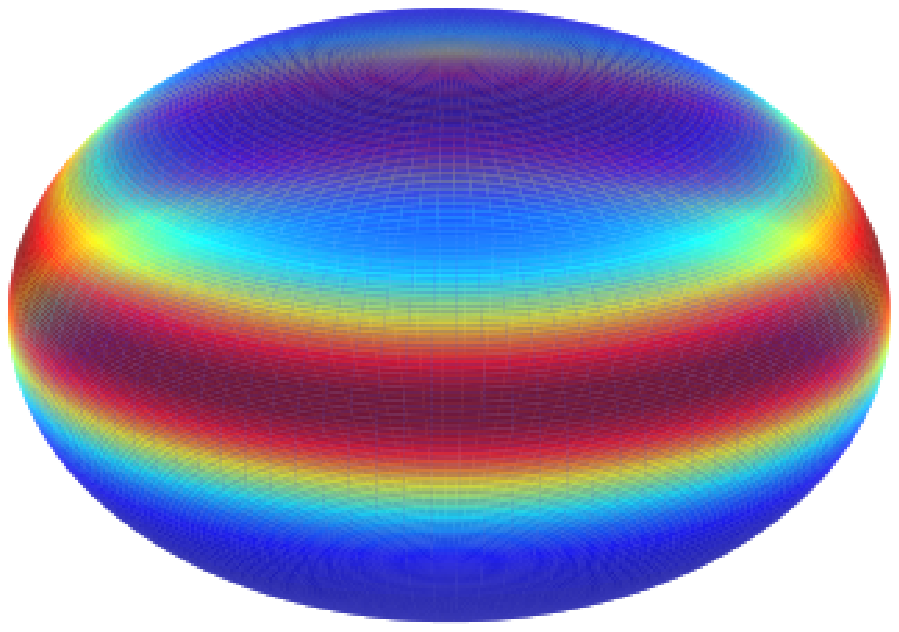}\\[-4ex]\hspace{0.7em}\normalsize{$\Sigma_2$}}        
        \hspace{-0.7cm}\shortstack{\includegraphics[scale = 0.25]{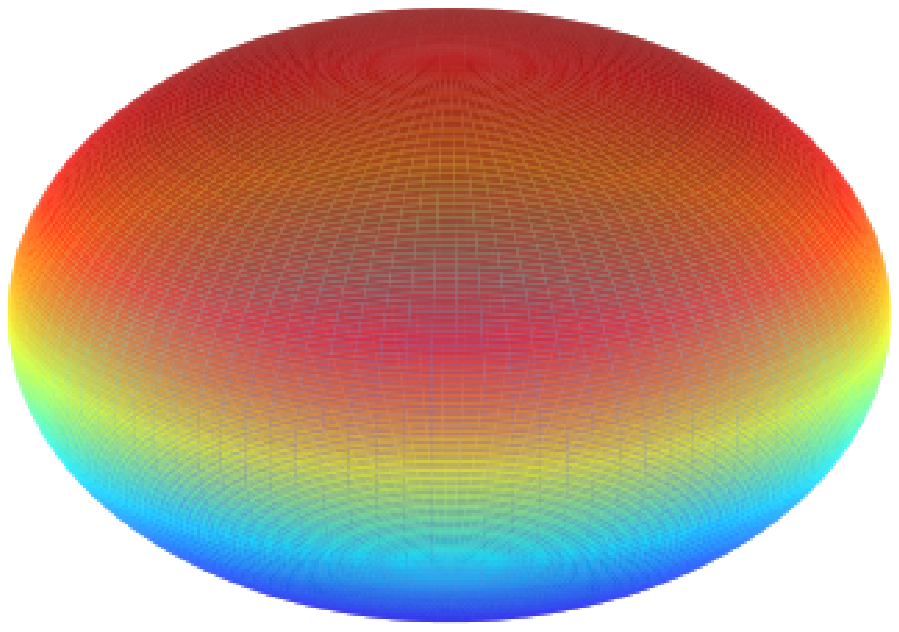}\\[-4ex]\hspace{0.7em}\normalsize{$\Sigma_3$}}
        \hspace{-0.7cm}\shortstack{\includegraphics[scale = 0.25]{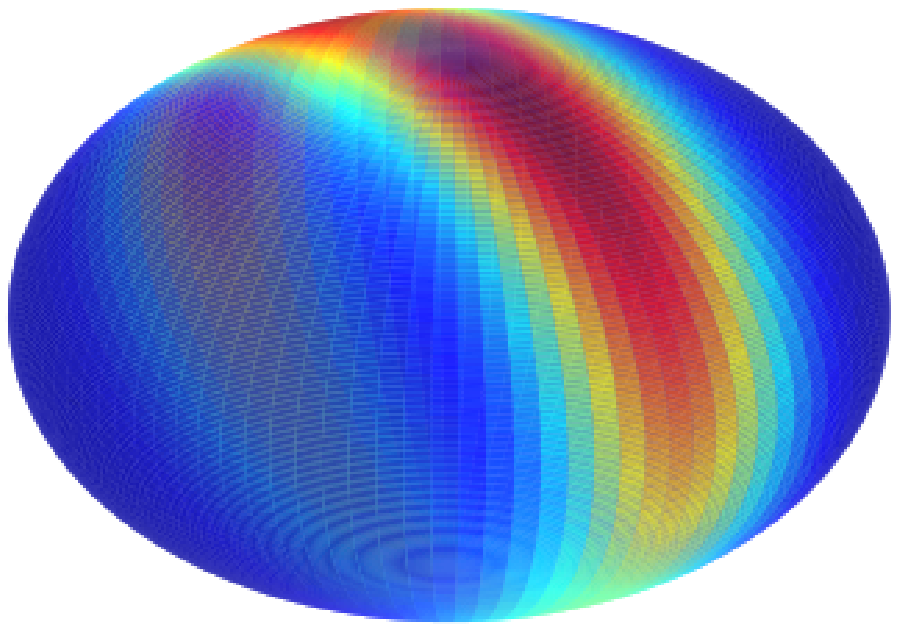}\\[-4ex]\hspace{0.7em}\normalsize{$\Sigma_4$}}
        \hspace{-0.7cm}\shortstack{\includegraphics[scale = 0.25]{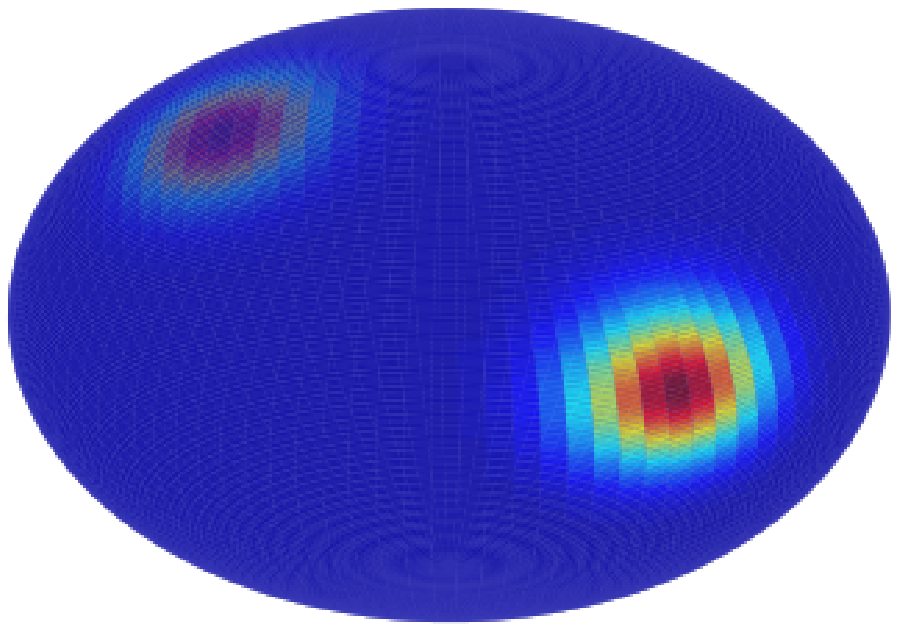}\\[-4ex]\hspace{0.7em}\normalsize{$\Sigma_5$}}       
        \\
        \subfigure[]{ \label{fig:q_T_phase_const_L}
		\includegraphics[scale = 0.45]{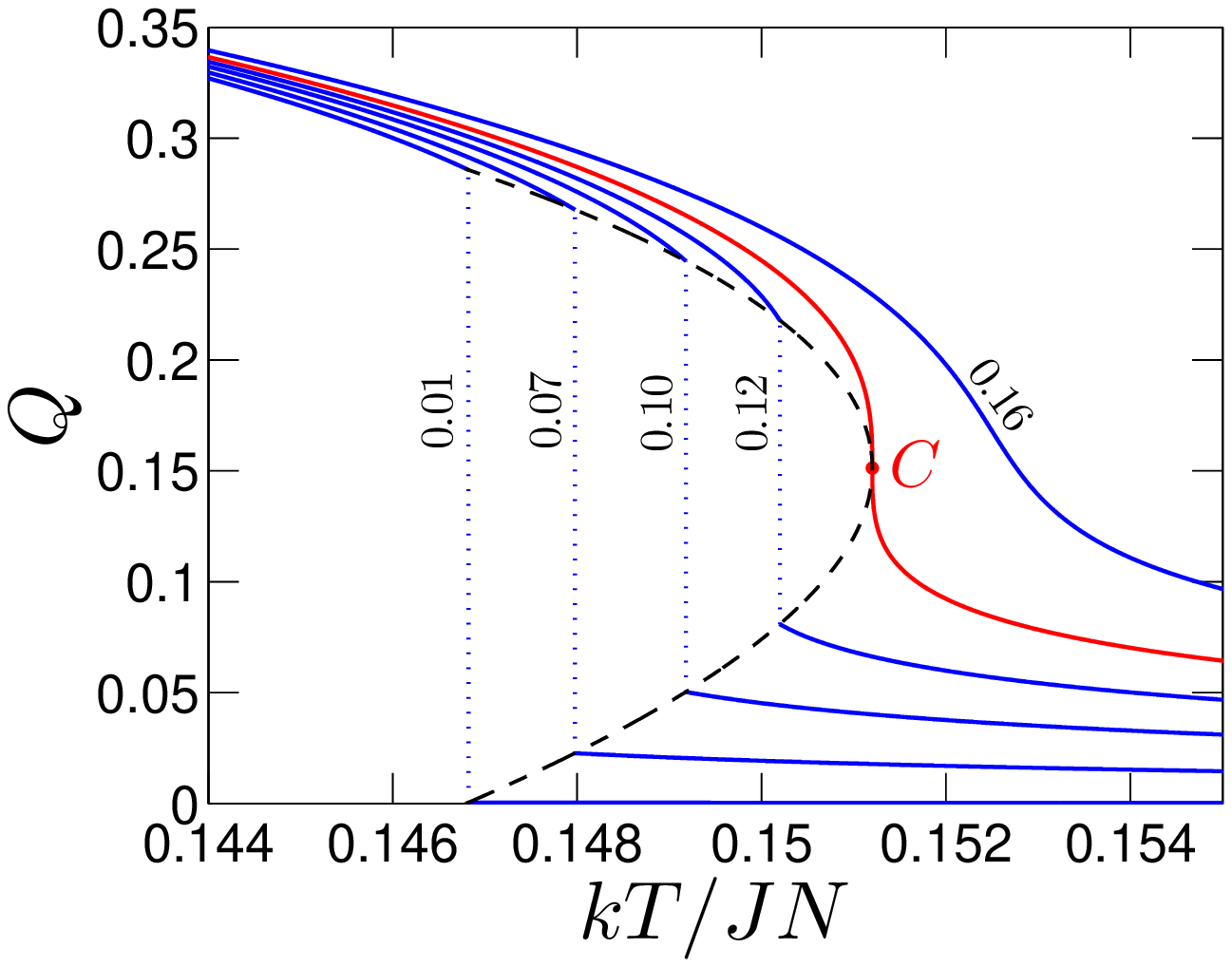} }
		\hspace{0.5in}
        \subfigure[]{ \label{fig:T_L_coexistence}
		\includegraphics[scale = 0.32]{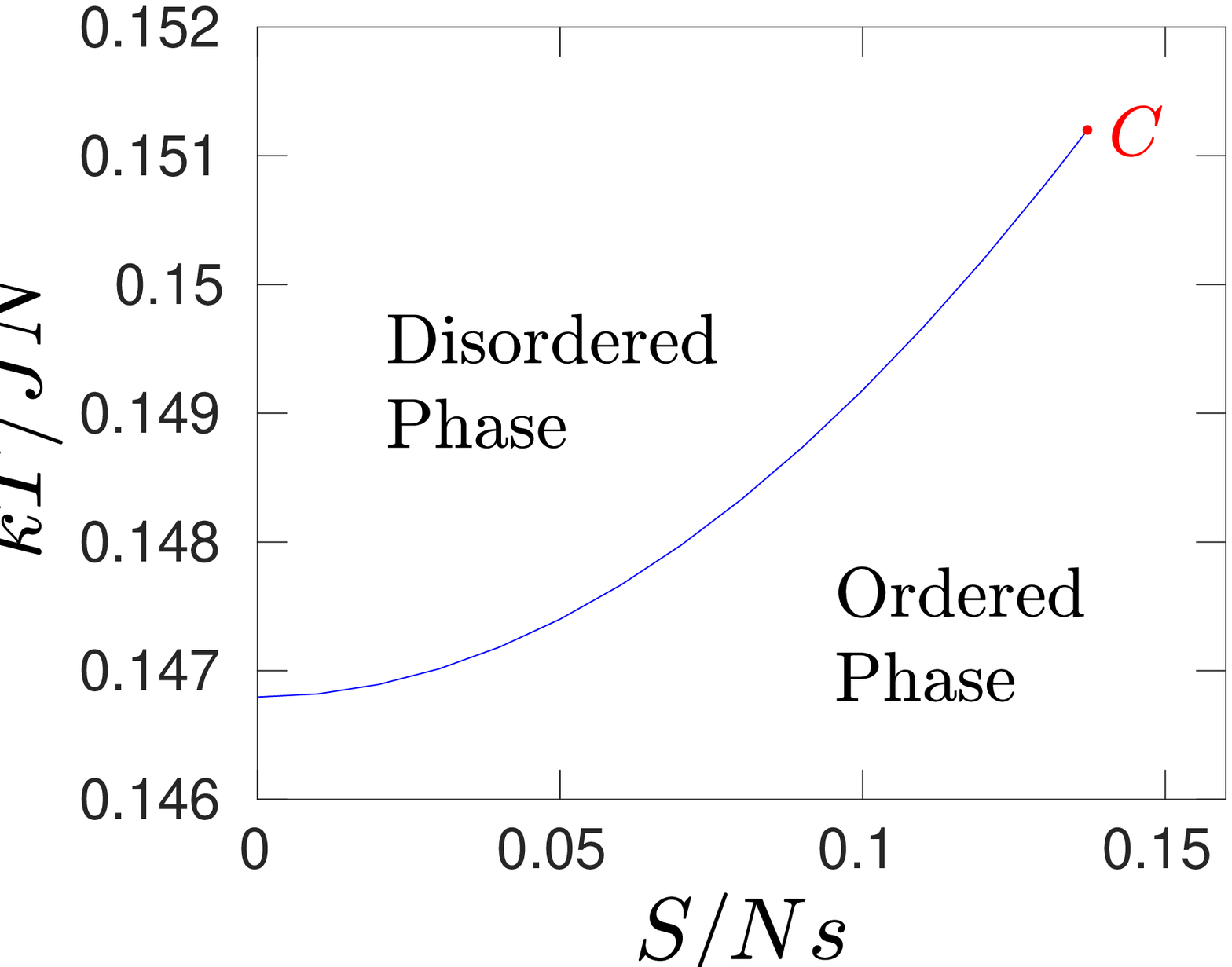} }
\caption{ The gravitational phase transitions of VRR are not affected by the kinetic energy term, even for disks with different moments of inertia. They are valid for the whole range of values $0 < \varepsilon < 1$, corresponding to $0 < mN/M < 1$.
   \textit{Top:} The distribution function of spin direction vectors $f_A(\theta,\phi)$ for the five, distinct non-equivalent thermodynamic equilibrium states that can be identified. $\Sigma_1$ represents the uniaxial, ordered phase, $\Sigma_2$ and $\Sigma_4$ are unstable states for any temperature and spin, $\Sigma_3$ represents the disordered phase, while $\Sigma_5$ represents a metastable biaxial phase which decays either to $\Sigma_1$ or $\Sigma_3$. 
   \textit{Bottom:}  (a) The order parameter $Q$, for $W=0$ with respect to temperature for various values of the total spin $\tilde{S}$. A first order phase transition occurs for $\tilde{S} < \tilde{S}_C$ between the ordered states $\Sigma_1$, with high $Q$, and disordered states $\Sigma_3$, with lower $Q$. It becomes second-order at the critical point $C$. 
    (b) The phase transition temperature as a function of the total spin.
    }
	\label{fig:F_q_w}
\end{center} 
\end{figure}

I introduce the $N$ matrices $\mathbf{q}_i$
\begin{equation}\label{eq:q_matrix}
	\mathbf{q}_i(\mathbf{n}_i) \equiv  \bm{n}_i \otimes \bm{n}_i - \frac{1}{3} \mathbf{I},\quad\text{that is}\quad
	q_{\mu\nu\,i} \equiv  n_{\mu,i}n_{\nu,i} - \frac{1}{3} \delta_{\mu\nu},
\end{equation}
where the Greek indices denote the coordinates of each disk. The normalized potential energy is then written as
\begin{equation}
	\tilde{U}(\bm{n}_i\cdot\bm{n}_j) \equiv \frac{U}{JN^2} = -\frac{1}{2N^2} \sum_{i=1}^N\sum_{j=1}^N \mathbf{q}_i(\bm{n}_i) \cdot \mathbf{q}_j(\bm{n}_j) + \frac{1}{3},
\end{equation}
where the dot denotes contraction $\mathbf{q}_i\cdot \mathbf{q}_j \equiv q_{\mu\nu,i}q_{\mu\nu,j}$ and I use the Einstein's summation rule for the Greek indices. 
The Gibbs partition function (\ref{eq:Xi}) becomes, incorporating also Eq. (\ref{eq:param_MF})
\begin{equation}\label{eq:Xi_order}
\Xi (\tilde{\beta},\tilde{\bm{\gamma}}_\text{S}) = \left(\frac{2 \pi \varepsilon h(\tilde{\beta})}{\tilde{\beta}}\right)^N\frac{e^{-N\tilde{\beta}/3}}{N!}\int \prod_{i=1}^N d\mathcal{A}_i \exp \left\lbrace -N\, \left( -\frac{\tilde{\beta}}{2}\frac{1}{N^2}\sum_{i=1}^N\sum_{j=1}^N \mathbf{q}_i(\bm{n}_i)\cdot \mathbf{q}_j(\bm{n}_j) + \tilde{\bm{\gamma}} \cdot\frac{1}{N} \sum_{i} \bm{n}_i\right) \right\rbrace,
\end{equation}
and $h(\theta_i)$ is given by Eqs. (\ref{eq:h_i}) subject to the assumption (
\ref{eq:param_MF}).

Following \cite{2017ApJ...842...90R}, I define the matrix $\mathbf{M}$ 
\begin{equation}
	\mathbf{M} = \frac{1}{N}\sum_{i=1}^N \mathbf{q}_i.
\end{equation}
It is 
\begin{equation}
	\frac{1}{N^2}\sum_{i=1}^N\sum_{j=1}^N \mathbf{q}_i \cdot \mathbf{q}_j = \mathbf{M}\cdot\mathbf{M}.
\end{equation}
I apply the Hubbard-Stratonovich transformation
\begin{equation}
	\exp \left(-\frac{1}{2} \mathbf{M}\cdot\mathbf{M}
	\right) = \left(\frac{1}{2\pi}\right)^\frac{9}{2} \int_{-\infty}^{+\infty} \prod_{\sigma,\lambda=1}^3 dQ_{\sigma\lambda} 
    \exp\left( \frac{1}{2} \mathbf{Q}\cdot\mathbf{Q}  - \mathbf{Q}\cdot \mathbf{M} \right),
\end{equation}
where I introduced the matrix $\mathbf{Q}$, which will play the role of the macroscopic variable that defines a macrostate, like the energy did in Eq. (\ref{eq:Xi_tilde}).
We get 
\begin{equation}
\Xi (\tilde{\beta},\tilde{\bm{\gamma}}) = \left( \frac{h(\tilde{\beta})}{\tilde{\beta}} \right)^N\left\lbrace\frac{e^{-N\tilde{\beta}/3}(2 \pi \varepsilon )^N}{(2\pi)^{9/2}N!} \right\rbrace
\int_{-\infty}^{+\infty} \prod_{\sigma,\lambda=1}^3 dQ_{\sigma\lambda} e^{-N \ln\left\lbrace
\int d\mathcal{A}\, \exp \left(- \tilde{\beta} (\frac{1}{2} \mathbf{Q}\cdot\mathbf{Q} - \mathbf{Q}\cdot \mathbf{q} ) - \tilde{\bm{\gamma}} \cdot \tilde{\bm{S}} \right)
\right\rbrace^{-1} }.
\end{equation}
The saddle point gives in the thermodynamic limit (\ref{eq:therm_limit}), $N\gg 1$
\begin{equation}\label{eq:G_order}
	\tilde{G}(\tilde{\beta},\tilde{\bm{\omega}}_\text{S}) = \inf_{\mathbf{Q}} \left\lbrace -\ln
\int d\mathcal{A}\, e^{- \tilde{\beta} (\frac{1}{2} \mathbf{Q}\cdot\mathbf{Q} - \mathbf{Q}\cdot \mathbf{q} ) - \tilde{\bm{\gamma}} \cdot \tilde{\bm{S}} } \right\rbrace 
+ g(\tilde{\beta}),
\end{equation}
where $g(\tilde{\beta})$ some function of $\tilde{\beta}$ which gives rise to a kinetic energy term. We will discuss this term in the next section (\ref{sec:K_VRR}).
The matrix $\mathbf{Q}$ encapsulates the ``order parameters'' of the system. Defining
\begin{equation}\label{eq:xi}
	\xi (\mathbf{Q}) = -\ln
\int d\mathcal{A}\, e^{- \tilde{\beta} ( \frac{1 }{2} \mathbf{Q}\cdot\mathbf{Q} - \mathbf{Q}\cdot \mathbf{q} ) - \tilde{\bm{\gamma}} \cdot \tilde{\bm{S}} }
\end{equation}
the condition (\ref{eq:G_order}) gives at equilibrium
\begin{equation}\label{eq:xi_cond}
	\left.\frac{\partial \xi}{\partial \mathbf{Q}}\right|_{\mathbf{Q}=\mathbf{Q}_{\text{e}}} = 0
\end{equation}
which results to the self-consistency equation for the matrix order parameter (I drop the subscript `e' here and imply equilibrium values for $\mathbf{Q}$)
\begin{equation}\label{eq:Q_self}
	\mathbf{Q} = \int d\mathcal{A}\,\mathbf{q}(\bm{n})\, f_A(\bm{n}),
\end{equation}
where 
\begin{equation}\label{eq:f_A_MF}
	f_A(\bm{n}) = \frac{ e^{\tilde{\beta} \mathbf{Q}\cdot \mathbf{q}(\bm{n}) - \tilde{\bm{\gamma}} \cdot \tilde{\bm{S}} (\bm{n}) } }{\int d\mathcal{A} \,e^{\tilde{\beta} \mathbf{Q}\cdot \mathbf{q}(\bm{n}) - \tilde{\bm{\gamma}} \cdot \tilde{\bm{S}} (\bm{n}) } },
\end{equation}
The self-consistency equations define the equilibrium configurations of the system and are equivalent to a mean field theory\footnote{The same result (\ref{eq:Q_self}) one may obtain by calculating the first order variations with respect to the mean-field entropy $\mathcal{S}_\text{MF} = -k \int d\tau f_\text{MF} \ln f_\text{MF}$ of one-particle distribution function $f_\text{MF}$, subject to the constraints of fixed energy, spin, and number of particles. One will get $f_\text{MF} = f_A$. I shall not perform this here. The interested reader may consult \cite{2017ApJ...842...90R} on the spacial distribution, while we will discuss the kinetic part in section \ref{sec:K_VRR}. }
They do not depend on the kinetic term. 

Even in case of disks with different moments of inertia $I_i$, we would get the same self-consistency equations and therefore identical spacial equilibrium distributions. This is evident from equation (\ref{eq:U_ens}) for the ensemble mean of $U$ on the $N$-particle distribution, which gives 
\begin{equation}
	\LA U\RA_\text{ens} = \frac{ \int \prod_{i=1}^N d\mathcal{A}_i\, U \,  e^{- ( \beta U + \bm{\gamma} \cdot \bm{S}) } }{\int \prod_{i=1}^N d\mathcal{A}_i\,  \,e^{- ( \beta U + \bm{\gamma} \cdot \bm{S} ) } }.
\end{equation}
The functions $h_i(\beta)$, defined in (\ref{eq:h_i}) and which emerge from the kinetic term $I_i\omega_i^2/2$, drop out in the ensemble mean of quantities depending only on spacial degrees of freedom. Thus, the kinetic term does not affect the spacial equilibrium distribution and as a consequence the later is not affected by the value of $\varepsilon$. The VRR gravitational phase transitions discovered for a zero kinetic term in \cite{2017ApJ...842...90R} do hold in any case.

The self-consistency equations (\ref{eq:Q_self}) have been solved in \cite{2017ApJ...842...90R}.
It has been proven there that the total spin is aligned with the eigenvalues of the matrix $\mathbf{Q}$. In the spherical coordinate system of sections \ref{sec:Dyn}, \ref{sec:Stat}, this matrix may be written as 
\begin{align}\label{eq:Qmatrix}
\mathbf{Q}
=\left(
\begin{array}{ccc}
	-\frac{1}{2}Q + \frac{1}{2}W & 0 & 0 \\
	0 &	-\frac{1}{2}Q-\frac{1}{2}W & 0 \\
	0 &	0 & Q
\end{array}
\right)
\end{align}
where we define
\begin{align}
\label{eq:Q_def}	Q &= \int q f_A(\theta,\phi)  d\mathcal{A}\, ,\quad
q = \cos^2\theta -\frac{1}{3} ,
\\
\label{eq:W_def}	W &= \int w f_A(\theta,\phi)  d\mathcal{A}\, ,\quad
w = \sin^2\theta \cos 2\phi .
\end{align}
The system is characterized by two order parameters $Q$ and $W$. The former describes deviation from spherical symmetry and the later from axial symmetry, so that $W=0$ corresponds to axially symmetric states and $Q=W=0$ to exactly isotropic states. 

The system is subject to first order \textit{gravitational} phase transitions between a uniaxial, ordered phase and a disordered phase, which become second order at a certain critical point occuring for spin $\tilde{S}_\text{C} = 0.14$ and temperature $\tilde{T}_\text{C} = 0.15$. 
In Figure \ref{fig:F_q_w} there are depicted the various phases of the system and the phase diagram for axially symmetric equilibria. 

\section{Precession and Nutation of Orbital Planes}\label{sec:K_VRR}

The kinetic energy of the system is associated with the angular velocity dispersion of the disks, about any of their diameters, as in Eq. (\ref{eq:K_resc}) of section \ref{sec:Bounds}, that I rewrite here as
\begin{equation}\label{eq:K_dispersion}
	K = \frac{\varepsilon}{2} JN^2 \left(\frac{1}{N}\sum_{i=1}^N \tilde{I}_i \omega_i^2\right).
\end{equation}
The angular velocity magnitude $\omega_i$, defined in (\ref{eq:omega_resc}), is the magnitude of nutation $\dot{\theta}$ and precession $\dot{\phi}$ of the effective disks (orbital planes) scaled with the characteristic timescale, which gives
\begin{equation}\label{eq:omega_tilde}
	\omega_i = \left\lbrace
	\left(\frac{d\theta_i}{d\tilde{t}}\right)^2
	+ \sin^2\theta_i \left(\frac{d\phi_i}{d\tilde{t}}\right)^2
	\right\rbrace^{1/2} = \sqrt{ \omega_{\text{nut},i}^2 + \omega_{\text{prec},i}^2 },
\end{equation} 
where I define the angular velocities describing nutation and precession
\begin{equation}
	\omega_{\text{nut},i} \equiv \frac{d\theta_i}{d\tilde{t}},\quad
	\omega_{\text{prec},i} \equiv \sin\theta_i \frac{d\phi_i}{d\tilde{t}}.	
\end{equation}
Note that while $\omega_i$ is strictly positive, being a magnitude, the $\omega_{\text{nut},i}$ and $\omega_{\text{prec},i}$ may be positive or negative, being rates of change.

Having reviewed this, we will in the followings inspect the expression (\ref{eq:K_VRR}) for the  kinetic energy as an ensemble average, which may be written as
\begin{equation}\label{eq:K_VRR-2}
	K_\text{ens} = NkT\left\lbrace 1 - \frac{\varepsilon\tilde{\beta}}{2}\frac{1}{N}\sum_{i = 1}^N \tilde{I}_i\,\frac{
	 e^{-\tilde{I}_i\frac{\varepsilon\tilde{\beta} }{2}} }{ 1 - e^{-\tilde{I}_i\frac{\varepsilon\tilde{\beta} }{2}}} \right\rbrace.
\end{equation}

\subsection{Identical Effective Disks}

Let us assume first that all effective disks acquire about the same moment of inertial $\tilde{I}_i \simeq 1$, which means $I_i \simeq I$, where $I=\sum_i I_i/N$ is the mean moment of inertia. Then, according to the expression of the moment of inertia Eq. (\ref{eq:I_inertia_app}), all bodies $m_i$ are of about the same mass, lie at about the same distance $a_i$ from the center of the cluster and have about the same eccentricities. The effect of the later is limited and cannot contribute a ratio greater than $I_i/I = 2.5$ or lower than $0.4$. 

The kinetic energy (\ref{eq:K_VRR-2}) becomes simply 
\begin{equation}\label{eq:K_I-eq}
	K_\text{ens} = Nk T\left(1 - 
	\frac{\varepsilon\tilde{\beta}}{2}\frac{1}{ e^{\frac{\varepsilon \tilde{\beta}}{2}}-1}\right) 
,\quad \Tilde{I}_i \simeq 1 ,\;\forall\, i.
\end{equation}
The kinetic energy depends on the ratio $\varepsilon\tilde{\beta} = \varepsilon\, JN/kT$. At very low values $kT/JN \ll \varepsilon$, temperature acquires its typical kinetic interpretation $K\sim NkT$, since the phase space boundary lies at infinity in this case. 

Equation (\ref{eq:K_I-eq}) reveals further that the kinetic energy is bounded 
\begin{equation}\label{eq:K_c}
	K_\text{ens} \leq  K_{c} \equiv \frac{\varepsilon}{4} JN^2, 
\end{equation}	
as expected by the VRR-bound in phase space (\ref{eq:p_constraint}). This upper limit is attained at sufficiently high temperature $kT/JN\gg \varepsilon$.

\begin{figure}[tbp]
\begin{center}
		\includegraphics[scale = 0.4]{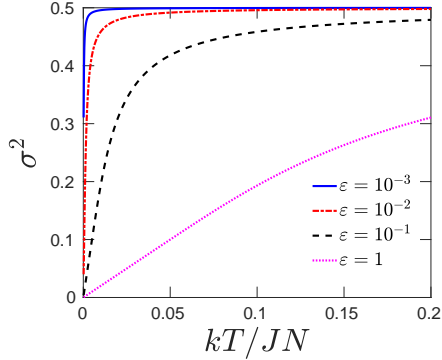} 
	\caption{ The velocity dispersion $\sigma^2 =  <\omega^2>$ of VRR, given in Eq. (\ref{eq:sigma_ens}), with respect to temperature $T$ for disks with equal moments of inertia $\tilde{I}_i=1$. It is the dispersion of precession and nutation angular velocity magntitude $\omega$ scaled by the characteristic frequency $t_c^{-1}$,  of effective disks (orbital planes) at thermal equilibrium. Four values of $\varepsilon \sim mN/(M_\star + M_\bullet)$ are considered. In case that the massive body $M_\bullet$ dominates completely the potential ($\varepsilon \lesssim 10^{-3}$), the equilibrium states attain the maximum possible dispersion $\sigma_\text{max}^2 = 1/2$ already at very low temperature ($kT/JN \lesssim 10^{-3}$) corresponding to very ordered states. This constant value is sustained for any higher temperature, including the disordered states. For $0.1 \lesssim \varepsilon \leq 1$, the temperature does affect the velocity dispersion at any temperature $kT/JN \leq kT_P/JN = 0.15$. I emphasize $T_P$, because at about this value occur all gravitational phase transitions.  Especially for $\varepsilon = 1$, at $T \leq T_P$, the temperature acquires a typical kinetic interpretation $kT \sim \sigma^2$. Such type of dependence is true at any temperature $kT/JN \ll \varepsilon$ for all $\varepsilon$ no matter how small. The angular velocity satisfies $\omega< 1$ in every case for all disks.
	\label{fig:sigma_T-epsilon}}
\end{center} 
\end{figure}

At thermal equilibrium, the dispersion
\begin{equation}
	\sigma^2 \equiv \frac{1}{N}\sum_i \omega_i^2,
\end{equation}
will acquire its ensemble average value
\begin{equation}\label{eq:sigma_ens}
	\sigma_\text{ens}^2 =  \LA \omega^2 \RA_\text{ens} = \frac{2K_\text{ens}}{\varepsilon \,JN^2} = 
	\frac{2}{\varepsilon\tilde{\beta}} \left(1 - 
	\frac{\varepsilon\tilde{\beta}}{2}\frac{1}{ e^{\frac{\varepsilon \tilde{\beta}}{2}}-1}\right) ,
\end{equation}
that is
\begin{equation}
	\sigma \rightarrow \sigma_\text{ens}(\varepsilon\tilde{\beta}), \quad \text{for }\;t \rightarrow t_\text{VRR} .
\end{equation}
This is a prediction regarding the dynamics of (\ref{eq:eom_VRR}) for any $\varepsilon$, provided they satisfy the constraint (\ref{eq:A_con}). Likewise it is a prediction about the $SO(3)$ dynamics (\ref{eq:eom_VRR_SO3}), which is an approximation of (\ref{eq:eom_VRR}) to zeroth order in $\varepsilon \ll 1$. 

In Figure (\ref{fig:sigma_T-epsilon}) is depicted the dispersion with respect to temperature for different values of $\varepsilon$. 
It is bounded
\begin{equation}\label{eq:sigma_c}
	\sigma_\text{ens} \leq \sigma_c	
,\quad
	\sigma_c = \frac{\sqrt{2}}{2}.
\end{equation}
The $\sigma_c$ does correspond to the characteristic energy (\ref{eq:K_c}).
This characteristic dispersion of VRR is attained in the limit $\varepsilon\tilde{\beta } \ll 1$, 
\begin{equation}
	\sigma \rightarrow \sigma_c ,\quad \text{for }\;  kT/JN \gg \varepsilon.
\end{equation}
For $\varepsilon \ll 1$, we get that, in effect, this is the dispersion for all equilibria above $T=0$!  The $SO(3)$ evolution (\ref{eq:eom_VRR_SO3}) will attain dispersion equal to $1/\sqrt{2}$ at thermal equilibrium. 

At this point I remark, that due to the long-range character of the interaction, the system may be trapped in quasi-stationary states. The system's evolution, described by Eq. (\ref{eq:eom_VRR}), is governed by a Vlasov equation. The system may be subject to violent relaxation that will result to a quasi-stationary state of the Lynden-Bell type \cite{LyndenBell:1966bi}, like in the case of Hamiltonian Mean Field Model \cite{2006EPJB...53..487C,2019JSMTE..04.3201G}. In such a state the nutation and precession of orbital planes will be much more intense. This possibility requires further investigation.

According to the partition function (\ref{eq:Z_can_om}), the one-particle distribution of angular velocity  is 
\begin{equation}
	f_K(\omega) = 
	\frac{\varepsilon \tilde{\beta}}{2\pi \, h(\varepsilon\tilde{\beta})} e^{-\frac{\tilde{\beta \varepsilon}}{2}\omega^2}
	\left\lbrace
		\begin{array}{ll}
			\simeq 1/\pi &, \text{ for } kT/JN \gg \varepsilon \\
			\simeq \frac{\varepsilon \tilde{\beta}}{2\pi } e^{-\frac{\tilde{\beta \varepsilon}}{2}\omega^2} &, \text{ for } kT/JN \ll \varepsilon
		\end{array}
	\right.
\end{equation}
Recall that $\omega \in [0,1]$, $u\in [0,2\pi]$ and that integration is performed in  $ \omega d\omega\, du$. The integration $\int_0^1 \omega^2 f_K(\omega) \,2\pi \omega d\omega$ will yield exactly the ensemble's dispersion (\ref{eq:sigma_ens}).

By use of (\ref{eq:nut}), (\ref{eq:prec}), we may calculate the ensemble averages of nutation and precession velocity. These are zero because of the odd cosine and sine terms. However, their dispersion is non-zero, and on the contrary we get
\begin{equation}
\sqrt{\LA \omega_\text{nut}^2\RA_\text{ens} } =
\sqrt{\LA \omega_\text{prec}^2 \RA_\text{ens}} = 
\frac{1}{2}\sigma_\text{ens}.
\end{equation}

\subsection{Different Families of Effective Disks}

\begin{figure}[tbp]
\begin{center}
		\includegraphics[scale = 0.4]{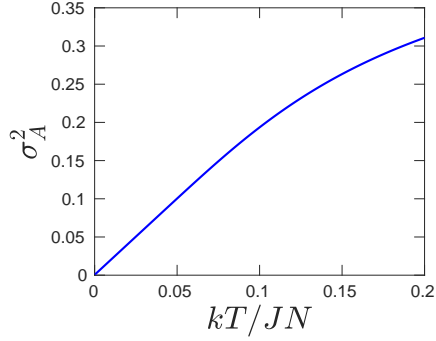} 
	\caption{ I consider $\varepsilon = 10^{-3}$ and two families of effective disks in VRR. The family $A$ with high moments of inertia $\tilde{I}_A = \varepsilon^{-1}$ and the family $B$ with low moments of inertia $\tilde{I}_B = \varepsilon$. It is depicted the quantity $< \tilde{I}\omega >$ which is nearly equal to $\sigma_A = \LA \omega_A^2\RA$ for these parameters' values. We infer that although $\varepsilon \ll 1$, the temperature may acquire a kinetic interpretation even up to the phase transition point $kT_P/JN =0.15$, for disks with sufficiently high moments of inertia with respect to the rest of the cluster of disks.
	\label{fig:sigma_families}}
\end{center} 
\end{figure}

Assume that there are $n\ll N$ different families of disks with different moments of inertia $\tilde{I}_{F ,i}$, for the $F_i$ family and the multitude of each family is $N_{F,i}$. The kinetic energy Eq. (\ref{eq:sigma_ens}) becomes
\begin{equation}\label{eq:K_fam}
	K_\text{ens} =  
	NkT \left(1 - 
	\frac{N_{F,1}}{N}\frac{\kappa_1}{ e^{\kappa_1}-1} - 	\cdots - \frac{N_{F,n}}{N}\frac{\kappa_n}{ e^{\kappa_n}-1} \right).
\end{equation}
It depends now on the quantities
\begin{equation}
	\kappa_i = \tilde{I}_{F,i}\frac{\varepsilon \tilde{\beta}}{2},\quad i = 1,\ldots,n,
\end{equation}
and the ratios $N_{F,i}/N$. 
The kinetic energy is now related to the quantity $<\tilde{I}\omega^2>$, instead of single dispersion and we have
\begin{equation}\label{eq:dispersion_fam}
	\frac{N_{F,1}}{N}\tilde{I}_{F,1} \sigma_{F,1}^2 + \cdots +\frac{N_{F,n}}{N}\tilde{I}_{F,n} \sigma_{F,n}^2  =  
	\frac{2}{\varepsilon\tilde{\beta}} \left(1 - 
	\frac{N_{F,1}}{N}\frac{\kappa_1}{ e^{\kappa_1}-1} - 	\cdots - \frac{N_{F,n}}{N}\frac{\kappa_n}{ e^{\kappa_n}-1} \right).
\end{equation}

Due to the dependence on the moments of inertia it can very well be the case that $\varepsilon \ll 1$, but $\kappa_{i} = \mathcal{O}(1)$ for a family $F_i$, if the moments of inertia deviate from the mean value by $\varepsilon^{-1}$. This is realistic. For example, consider $\varepsilon = 10^{-3}$, which may refer to a subcluster of $10^4$ stars of mass $M_\odot$ and a central massive black hole $10^7M_\odot$. Then a family $A$ of stars that may be $\sqrt{10^3} \simeq 30$ times further away than the rest of stars satisfies the relation $\tilde{I}_A \varepsilon = 1$. For demonstrating purposes, in Figure \ref{fig:sigma_families} I consider such a system, assuming also $\tilde{I}_B = \varepsilon$.
The number of disks are then $N_A/N = (1-\tilde{I}_B)/(\tilde{I}_A-\tilde{I}_B)$, $N_B/N = (\tilde{I}_A - 1)/(\tilde{I}_A-\tilde{I}_B)$. It is therefore $\tilde{I}_B N_B/N \ll 1$, $\tilde{I}_B \varepsilon\ll 1$, while the corresponding quantities for $A$ are $\simeq 1$. 
I plot the quantity
\begin{equation}
		\LA \tilde{I}\omega^2\RA  =  
	\frac{2}{\varepsilon\tilde{\beta}} \left(1 - 
	\frac{N_A}{N}\frac{\tilde{I}_{A}\frac{\varepsilon \tilde{\beta}}{2}}{ e^{\tilde{I}_{A}\frac{\varepsilon \tilde{\beta}}{2}}-1} - \frac{N_{B}}{N}\frac{\tilde{I}_{B}\frac{\varepsilon \tilde{\beta}}{2}}{ e^{\tilde{I}_{B}\frac{\varepsilon \tilde{\beta}}{2}}-1} \right).
\end{equation}
It is 
\begin{equation}
			\LA \tilde{I}\omega^2\RA  =  \frac{N_A}{N}\tilde{I}_A\sigma_A^2 + \frac{N_B}{N}\tilde{I}_B\sigma_B^2 \simeq \sigma_A^2 .
\end{equation}
Even though it is $\varepsilon = 10^{-3} \ll 1$, the dispersion of the family $A$, with $\tilde{I}_A = \varepsilon^{-1}$ follows a linear dependence on temperature almost up to the phase transition temperature $kT_P/JN = 0.15$.
In general, disks with higher moment of inertia acquire less velocity dispersion, which depends on temperature up to higher values of the later.

\section{Continuous Rigid Bodies}\label{sec:RB}

The case of continuous rigid bodies is much simpler than effective rigid disks, whose rigidity emerges at timescales $t>t_c$. The continuous rigid bodies are not subject to the phase space boundary constraint (\ref{eq:p_constraint}). The integration of the partition function is calculated in the whole range $\tilde{p} \in (-\infty, +\infty) $ giving
\begin{align}
	Z_\text{RB} &= \frac{1}{N!}\int_0^\pi\int_0^{2\pi}\iint_{-\infty}^{+\infty} \prod_{i=1}^N d\theta_i\, d\phi_i\, 
	\left\lbrace d\tilde{p}_{\theta,i}\, d\tilde{p}_{\phi,i} \,  
	  e^{- \tilde{\beta}\sum \frac{1}{2\varepsilon } \left( \frac{\tilde{p}_{\theta,i}^2}{\tilde{I}_i}  + \frac{\left(\tilde{p}_{\phi,i} - \tilde{s}_i \cos\theta_i \right)^2}{ \tilde{I}_i\sin^2\theta_i} \right) }
	  \right\rbrace 
	e^{- \beta \,U } \delta (\bm{S} - \sum_i s_i \bm{n}_i),\\
	\label{eq:Z_can_RB}
	&= \frac{1}{N!}
	\int_0^\pi \int_0^{2\pi}
	\prod_{i=1}^N d\mathcal{A}_i\,
	\left\lbrace \int_0^{2\pi} du_i \int_0^
	\infty d\omega_i \,\varepsilon^2 \tilde{I}_i^2\,\omega_i\,
	  e^{-\sum \frac{ \tilde{\beta} \varepsilon}{2} \tilde{I}_i\omega_i^2} 
	  \right\rbrace 
	e^{- \beta\,U }\delta (\bm{S} - \sum_i s_i \bm{n}_i),
\end{align}
which gives
\begin{equation}
		Z_\text{RB} = \frac{1}{N!}
		\left(\frac{2\pi\varepsilon}{\beta}\right)^N
	\int_0^\pi \int_0^{2\pi}
	\prod_{i=1}^N d\mathcal{A}_i\,\tilde{I}_i	e^{- \beta\,U }\delta (\bm{S} - \sum_i s_i \bm{n}_i).
\end{equation}
The partition function of the Gibbs-canonical ensemble likewise is written as
\begin{equation}
		\Xi_\text{RB} = \frac{1}{N!}
		\left(\frac{2\pi\varepsilon}{\beta}\right)^N
	\int_0^\pi \int_0^{2\pi}
	\prod_{i=1}^N d\mathcal{A}_i\,\tilde{I}_i	e^{- \beta\,U -\tilde{\bm{\gamma}} \cdot \bm{S} }.
\end{equation}
The kinetic energy is simply
\begin{equation}
	K_\text{RB} = NkT.
\end{equation}
The ensemble mean of potential energy 
\begin{equation}
		U_\text{RB} = 
\frac{\int \prod_{i=1}^N d\mathcal{A}_i\,	U \, e^{- \beta\,U -\tilde{\bm{\gamma}} \cdot \bm{S} }}{ \int \prod_{i=1}^N d\mathcal{A}_i\,	e^{- \beta\,U -\tilde{\bm{\gamma}} \cdot \bm{S} } },
\end{equation}
in the case of equal couplings and spins will give the same self-consistency equations with VRR, Eq. (\ref{eq:Q_self}). Therefore the system is subject to the same phase transitions described in \ref{sec:GPT}.

\section{Conclusions}\label{sec:conclusions}

I argue here that the general, VRR Hamiltonian in the time-averaging framework is Eq. (\ref{eq:H_VRR_sph}). It is directly analogous to rigid-body dynamics. It is a function of canonical variables, namely the Euler angles and their generalized momenta. The general, VRR equations of motion are Eqs. (\ref{eq:eom_VRR}), subject to the constraint (\ref{eq:A_con}). The following identifications $\varepsilon \equiv IJN/s^2 \sim mN/M$ and $t_c \equiv s/JN \sim t_\text{aps}$ emerge naturally in the canonical dynamics. They connect properties of the implicit system --orbiting point masses-- on the one hand with the effective system --rigid annular disks-- on the other. 

The time-averaging imposes boundaries on the canonical generalized momenta of the resulting canonical phase space, Eq. (\ref{eq:p_constraint}).  
The study of statistical mechanics induced by the effective Hamiltonian on this bounded phase space gives the partition functions (\ref{eq:Z_can}), (\ref{eq:Xi}) and the thermodynamic limit (\ref{eq:therm_limit}). 
The thermal equilibrium states are a result of the relaxation of spins' directions (direction vector of spin of the effective disks), identified with orbital planes' orientations. 

I validate the VRR \textit{gravitational} phase transitions and suggest their generalization to non-zero values of $\varepsilon$. These phase transitions occur between ordered phases, at low temperature, and disordered phases, at higher temperature. I emphasize that the phase transitions are purely gravitational, because no other effect or interaction besides gravity intervenes unlike for example the case of phase transitions in gravitational systems related to the presence of fermions or bosons \cite{1971CMaPh..24...22H,1986PhRvL..57.2485C}. The gravitational phases encountered here are manifestation solely of the gravitational interaction, averaged out towards an effective description, in the same sense that magnetic phase transitions are manifestation of the magnetic interaction or the liquid-gas transitions are manifestation of effective electromagnetic interactions. 

The dependence of spins' angular velocity dispersion on temperature is given in(\ref{eq:sigma_ens}) in the case of a family of bodies with equal moments of inertia. The dispersion depends on the quantity $I\varepsilon\beta$, and is also bounded by a characteristic VRR dispersion's value $\sigma_c$, Eq. (\ref{eq:sigma_c}). The boundary value $\sigma_c$ is attained in the limit $I\varepsilon\beta\rightarrow 0$.  For very small $\varepsilon$ values (dominating central massive object), even equilibria of very ordered states (low $T$) acquire the $\sigma_c$ value, which persists at any higher temperature. However, there can always be found a temperature low enough (therefore an equilibrium ordered enough) such that the boundary lies at infinity. This means in effect that for these states the dispersion (squared) follows a linear dependence on temperature. For $\varepsilon \gtrsim 10^{-1}$ the dispersion does not acquire the boundary value, but does depend on temperature $T$, for any $T\lesssim T_P$, where $T_P$ is the phase transition temperature. In addition, because of the dependence of the dispersion on the moment of inertia, different families of bodies acquire different dispersions on the same temperature. 

I remark that due to the long-range character of the interaction, the system may be trapped in quasi-stationary states. 
Just like in the case of Hamiltonian Mean Field Model \cite{2006EPJB...53..487C,2019JSMTE..04.3201G} the system may be subject to violent relaxation that will result to a quasi-stationary state of the Lynden-Bell type \cite{LyndenBell:1966bi}. In such a state the nutation and precession of orbital planes will be different than the one described here, although the spacial distribution will be the same. This possibility requires further investigation.

There are many directions in which this analysis can be improved in the future. I can suggest two of them, that will allow for the results to be more realistically applicable to physical systems. The first is the generalization of the phase transitions for non-equal couplings. The second is the generalization to the case of families of objects with different moments of inertia that will allow for more general results be drawn with respect to the dispersion of nutation and precession. 

\section*{Aknowledgements}

I am grateful to Scott Tremaine.

\appendix

\section{Keplerian Orbits and Apsidal Precession}\label{app:Keplerian}

In this Appendix I review well known material regarding Keplerian orbits and discuss apsidal precession.

A Keplerian bound orbit is an ellipse generated by the evolution of a gravitationally bound, 2-body system consisted of a body with mass $m$ and another body $M$, which interact  mutually via Newtonian gravitation. Here I assume that $M \gg m$ (in our $N$-body system, the analogue is the $i$th body with mass $m_i$ interacting with the central mass $M_i = M_\bullet + M_{\star,i}$, with $M_{\star,i} = M_\star (r<r_i)$). Due to angular momentum conservation the orbit lies on a plane. 

Consider a barycentric coordinate system, where I denote $r$ the distance between the two bodies and $\psi$ the angle between the position of $m$ and a reference direction on the same plane, with origin the barycentre. The 2-body Hamiltonian may be written in this system as
$
	H_\text{K} = (1/2)\mu (\dot{r}^2 + r^2\dot{\psi}^2) - G\mu M_\text{tot}/r
$,
where $M_\text{tot} = M + m \simeq M$ and $\mu = mM/M_\text{tot} \simeq m$ and therefore
\begin{equation}
	H_\text{K} = \frac{1}{2}m \left(\dot{r}^2 + r^2\dot{\psi}^2\right) - \frac{k}{r},\quad
	k = GmM.
\end{equation} 
The subscript ``K'' accounts for ``Keplerian''. The angular momentum conservation
\begin{equation}\label{eq:L_K}
	L_K = mr^2\dot{\psi}
\end{equation}
allows for elimination of $\psi$ and introduction of the effective spherical potential
\begin{equation}
	U_\text{eff}(r) = \frac{L_K^2}{2m r^2} - \frac{k}{r}.
\end{equation}
The Hamiltonian may now be decomposed as 
\begin{equation}
	H_K(r,\dot{r}) = \frac{1}{2}m\dot{r}^2 + U_\text{eff}(r) = E_K = \text{const}.
\end{equation}
Since $2(E-U_\text{eff}(r))/m = \dot{r}^2 \geq 0$,  the orbit is bound by two values 
\begin{equation}
	r_p \leq r \leq r_a
\end{equation}
called periastron ($r_p$) and apoastron ($r_a$) defined by
\begin{equation}
	E_K - U_\text{eff}(r_{p,a}) = 0.
\end{equation}
We get the solutions
\begin{equation}\label{eq:r_pa_K}
	r_p = a(1 - e),\;
	r_a = a(1 + e),\;	
\end{equation}
where the constants $a$, $e$ are defined by
\begin{equation}\label{eq:ae_K}
	a \equiv -\frac{k}{2E_K},\quad
	e \equiv \sqrt{1 + \frac{2E_K L_K^2}{mk^2}}.
\end{equation}
We will need several times in the followings the equations
\begin{equation}\label{eq:EU_eff}
	E_K - U_\text{eff} = -\frac{E_K}{r^2}(r_a - r)(r-r_p) = \frac{mk^2}{2L_K^2}\left( e^2 - x^2\right),\quad x = \frac{L_K^2}{mkr} - 1 
\end{equation}
The energy constraint gives
\begin{equation}\label{eq:E_K_dr}
	\frac{dr}{dt} = \pm\frac{1}{r} \sqrt{\frac{k}{ma}(r_a - r)(r - r_p)} = \pm \frac{k}{L_k}\sqrt{e^2 - x^2}.
\end{equation}
Combining with Eq. (\ref{eq:L_K}) in the form
\begin{equation}
	\frac{d\psi}{dt} = \frac{L_K}{mr^2},
\end{equation}
we get the differential equation of the orbit
\begin{equation}\label{eq:dpsi_K}
	d\psi = \pm \frac{L_K}{\sqrt{2m}}\frac{r^{-2}}{\sqrt{E_K - U_\text{eff}(r)}} dr
	 = \pm \sqrt{\frac{L_K^2 a}{km}} \frac{r^{-1}}{\sqrt{(r_a-r)(r-r_p)}} dr
	  = \mp \frac{1}{\sqrt{e^2 - x^2}} dx.
\end{equation}
This may be integrated straightforwardly by use of the integral
\begin{equation}\label{eq:int_arcsin}
	\int \frac{dx}{\sqrt{e^2 - x^2}} = \arcsin\left(\frac{x}{e}\right)
\end{equation}
and get
\begin{equation}\label{eq:r_K}
	r(\psi) = \frac{a(1-e^2)}{1+e \cos\psi},
\end{equation}
where we used (\ref{eq:ae_K}).
This is the equation of an ellipse with semi-major axis $a$ and eccentricity $e$.
We assumed the initial condition $r(\psi=0) = r_p$, equivalently $x(0) = e$. Equation (\ref{eq:ae_K}) gives the Keplerian energy and angular momentum of the system with respect to the semi-major axis and eccentricity as
\begin{align}
\label{eq:E_Kepler}
	E_K &= -G\frac{mM}{2a} \\ 
	L_K &= m\sqrt{GMa (1-e^2)}.
\end{align}
The binding energy of the binary is equal to $E_b = |E_K|$. 

We may define the orbital period $t_\text{orb}$ as the time needed to return to $r_p$ starting from this point. Using Eq. (\ref{eq:E_K_dr}) we can calculate  the period as
\begin{equation}
	t_\text{orb} = 2\sqrt{\frac{ma}{k}}\int_{r_p}^{r_a} \frac{r}{\sqrt{(r_a-r)(r-r_p)}}dr
	= 2\sqrt{\frac{a}{GM}}\int_{0}^{r_a-r_p} \frac{r+r_p}{\sqrt{r((r_a-r_p)-r)}}dr.
\end{equation}
The integrals
\begin{equation}
	\mathcal{I}_n(A) = \int_0^{A} \frac{r^n}{\sqrt{r(A-r)}}dr
\end{equation}
are related to the beta function
\begin{equation}\label{eq:integral_n}
	\mathcal{I}_n(A) = A^n B(n+\textstyle{\frac{1}{2},\frac{1}{2}}) =
	\displaystyle A^n \frac{\Gamma(\frac{1}{2})\Gamma(n+\frac{1}{2})}{\Gamma(n+1)} = \pi \frac{A^n}{4^n}\frac{(2n)!}{(n!)^2},
\end{equation}
where $\Gamma$ is the gamma function. It is now straightforward to get
\begin{equation}
	t_\text{orb} = 2\sqrt{\frac{a}{GM}}\left( \mathcal{I}_1(2ea) + a(1-e)\mathcal{I}_0(2ea)\right) 
\end{equation}
which gives for Keplerian orbits
\begin{equation}
	t_\text{orb,K} = 2\pi \sqrt{\frac{a^3}{GM}}.
\end{equation}

The line connecting $r_p$ (periapsis) and $r_a$ (apoapsis) is called the line of apsides. During one orbital period the line of apsides is dislocated by an angle $\Delta \psi_\text{aps} = 2\pi - \psi_\text{aps}(t_\text{orb})$ that may be calculated by use of equations (\ref{eq:dpsi_K}) as 
\begin{equation}\label{eq:psi_aps}
	\Delta\psi_\text{aps} = 2\pi - 2 \frac{L_K}{\sqrt{2m}}\int_{r_p}^{r_a}\frac{r^{-2}}{\sqrt{E_K - U_\text{eff}(r)}} dr.
\end{equation}
For Keplerian orbits we get straightforwardly by use of the integral (\ref{eq:int_arcsin}) that
\begin{equation}	
	\Delta\psi_\text{aps,K}  = 2\pi - 2 \int_{-e}^{+e}\frac{1}{\sqrt{e^2 - x^2}} dx = 0.
\end{equation}

In case that $\Delta\psi_\text{aps}  \neq 0$ we say that the orbit is subject to apsidal precession. The apsidal precession period is then
\begin{equation}
	t_\text{aps} = \frac{2\pi}{\Delta\psi_\text{aps}}t_\text{orb}.
\end{equation}
We are interested in the apsidal precession induced to an orbit by the collective effect of the (Newtonian)  gravitation from other bodies of the cluster. Let us assume that this effect amounts to an external effective spherical potential, which mimics the effect of an additional constant angular momentum  induced externally to the system
\begin{equation}
	U_\text{ext} = \frac{L_\text{ext}^2}{2mr^2} 
\end{equation}
and therefore acts as an additional centrifugal force
\begin{equation}
	F_\text{ext}(r) = -\frac{d}{dt}\left(\frac{L_\text{ext}^2}{2mr^2}\right) = \frac{L_\text{ext}^2}{mr^3}.
\end{equation}
The angular momentum of the body is preserved
\begin{equation}
	L = m r^2 \dot{\psi} = \text{const.}
\end{equation}
and the Hamiltonian may be written as
\begin{equation}
	H = \frac{1}{2}m\dot{r}^2 + U_\text{eff}(r) = E = \text{const.},
\end{equation}
where now
\begin{equation}
	U_\text{eff}(r) = \frac{L^2(1+\eta^2)}{2mr^2} - \frac{k}{r},
\end{equation}
and
\begin{equation}
	\eta = \frac{L_\text{ext}}{L}.
\end{equation}
In direct analogy with the Keplerian orbit, we get that the orbit is bounded due to $E - U_\text{eff}(r) \geq 0$ which gives now
\begin{equation}
	r_{p,\text{R}} \leq r \leq r_{a,\text{R}},
\end{equation}
with
\begin{equation}\label{eq:r_ap_aps}
	r_{p,\text{R}} = a_\text{R} (1 - e_\text{R}),\quad
	r_{a,\text{R}} = a_\text{R} (1 + e_\text{R}),
\end{equation}
where `R' stands for ``Rosette'' for reasons to be understood later and
\begin{equation}\label{eq:ae_aps}
	a_\text{R} \equiv -\frac{k}{2E},\quad
	e_\text{R} \equiv \sqrt{1 + \frac{2E L^2}{mk^2}(1+\eta^2)}.
\end{equation}
The differential equation of the orbit is now
\begin{equation}\label{eq:dpsi_aps}
	d\psi = \pm \frac{L}{\sqrt{2m}}\frac{r^{-2}}{\sqrt{E - U_\text{eff}(r)}} dr
	 = \pm \sqrt{\frac{L^2 a_\text{R}}{km}} \frac{r^{-1}}{\sqrt{(r_{a,\text{R}}-r)(r-r_{p,\text{R}})}} dr
	  = \mp \frac{1}{\sqrt{1+\eta^2}} \frac{1}{\sqrt{e^2 - x^2}} dx.
\end{equation}
Just like in the Keplerian case this equation can be integrated to give the equation of the orbit
\begin{equation}\label{eq:r_aps}
		r(\psi) = \frac{a_\text{R}(1-e_\text{R}^2)}{1+e_\text{R} \cos\left( H\psi\right)},
\quad
	H = \sqrt{1+\eta^2}.
\end{equation}
This equation describes an in-plane precessing orbit, which forms the shape of a rosette. The energy and angular momentum of the body are
\begin{align}
	E &= -G\frac{mM}{2a_R} \\ 
	L &= m\sqrt{GMa_R (1-e_R^2)}.
\end{align}

\begin{figure}[tbp]
\begin{center}
        \subfigure[$\frac{t_\text{aps}}{t_\text{orb}} = 5$, $\Delta t = 10 t_\text{aps}$]{ \label{fig:Rosette_tau-5}
		\includegraphics[scale = 0.35]{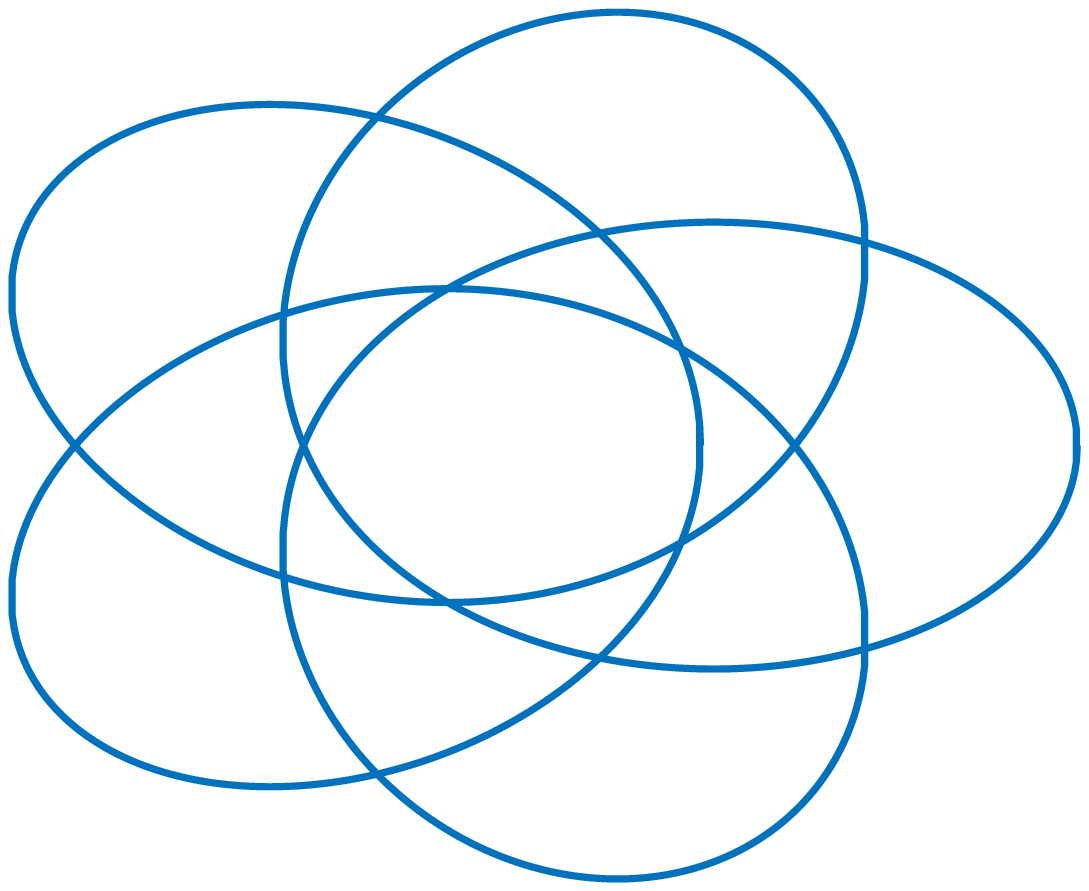} }
        \subfigure[$\frac{t_\text{aps}}{t_\text{orb}} = \sqrt{28}$, $\Delta t = 10 t_\text{aps}$]{ \label{fig:Rosette_tau-irr}
		\includegraphics[scale = 0.35]{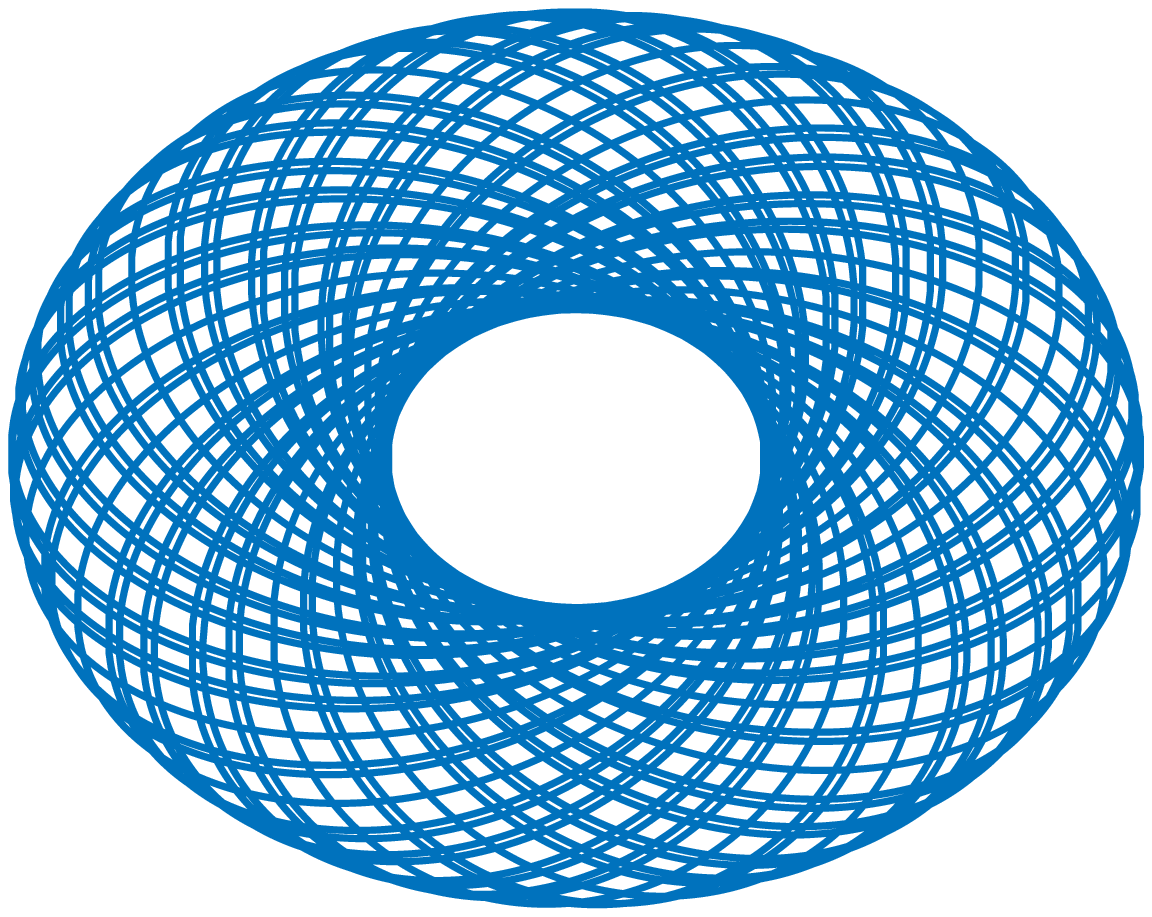} }
        \subfigure[$\frac{t_\text{aps}}{t_\text{orb}} = 100$, $\Delta t = t_\text{aps}$]{ \label{fig:Rosette_tau-100}
		\includegraphics[scale = 0.35]{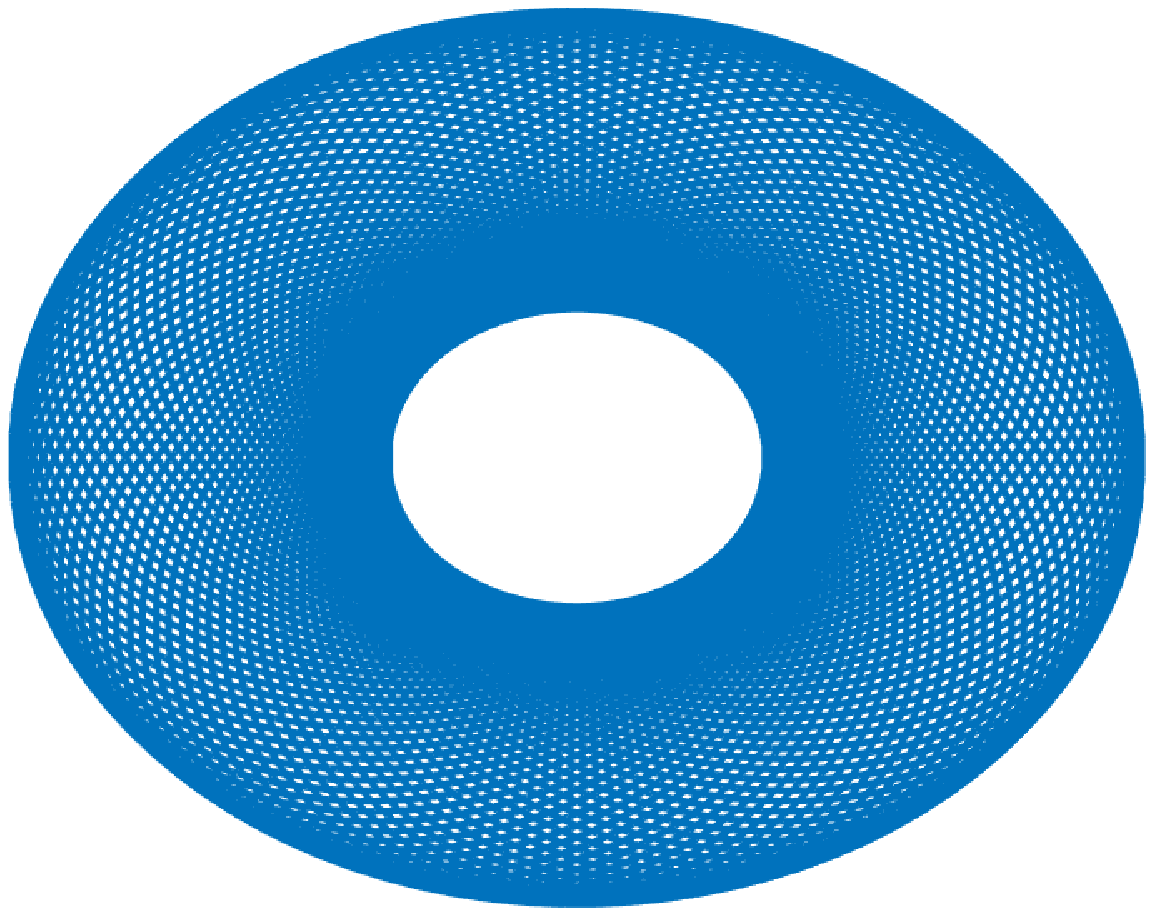} }
	\caption{The orbit is closed if $t_\text{aps}/t_\text{orb}$ is a rational number (a). If it is irrational then after a few precession periods the orbit resembles an annular disk as in (b). In (c) we set the apsidal precession period to be much longer than the orbital period and demonstrate that the resulting orbit resembles an annular disk within a single precession period. In all cases we set $a_\text{R}(1-e_\text{R}^2)=1$, $e_\text{R} = 0.5$.
		\label{fig:rosette}}
\end{center} 
\end{figure}

The orbital period is equal to that of a Keplerian orbit with semi-major axis $a_\text{R}$
\begin{equation}
	t_\text{orb} = 2\pi\sqrt{\frac{a_\text{R}^3}{GM}}.
\end{equation}
Let us calculate the period of precession $t_\text{aps}$. We have by Eq. (\ref{eq:psi_aps})
\begin{equation}
	\Delta\psi_\text{aps} = 2\pi - 2 \frac{1}{\sqrt{1+\eta^2}}\int_{-e}^{+e}\frac{1}{\sqrt{e^2 - x^2}} dx,
\end{equation}
which gives
\begin{equation}
	\Delta\psi_\text{aps} = 2\pi \left(1 - \frac{1}{\sqrt{1+\eta^2}} \right).
\end{equation}
Therefore 
\begin{equation}
	t_\text{aps} = \left(1 - \frac{1}{\sqrt{1+\eta^2}} \right)^{-1} t_\text{orb},
\end{equation}
which allows us also to express $H$ as
\begin{equation}
	H = \left(1 - \frac{t_\text{orb}}{t_\text{aps}} \right)^{-1}.
\end{equation}
The orbit defined by Eq. (\ref{eq:r_ap_aps}) is a rosette. If the ratio $(t_\text{aps}/t_\text{orb})$ is a rational number then the rosette closes after time interval equal to $t_\text{aps}$ as in Figure \ref{fig:Rosette_tau-5}. In the overwhelmingly more probably case that  $(t_\text{aps}/t_\text{orb})$ is irrational the rosette is non-closing and after a few apsidal precession periods the system resembles an annular disk as in Figure \ref{fig:Rosette_tau-irr}. In Figure \ref{fig:Rosette_tau-100} I show that if $t_\text{aps} \ll t_\text{orb}$, then within one precession period the orbit resembles an annular disk.

Note that the apsidal precession angular velocity is not constant in time, but equals
\begin{equation}
	\omega_\text{aps}(t) = \eta \frac{L}{m r(t)^2}.
\end{equation}
It represents the angular velocity of the special rotating frame of reference at which the orbit would appear as a Keplerian ellipse.

\section{Time-average of Kinetic Terms}\label{app:av}

In this Appendix we calculate the time-average of the kinetic terms (\ref{eq:kinetic_normal_def}), (\ref{eq:kinetic_planar_def}) and binary term \ref{eq:U_B_def}) in the case $M_\bullet \gg mN$, which refers to nearly-Keplerian orbits with $t_\text{orb} \ll t_\text{aps}$.  Since the apsidal precession proceeds very slowly we can integrate independently first over $t_\text{orb}$ and then over $t_\text{aps}$. 

We first calculate the normal term $K_\perp$. We assume that $v_{\perp,i} = \omega_\perp r$ and assume that $\omega_\perp$ consider variations of $\omega_\perp$ at timescales longer than the averaging timescales. We have
\begin{equation}\label{eq:K_int_2}
	K_\text{VR} = \sum_i \frac{1}{2} m_i \omega_{\perp,i}^2 \,\left( \frac{1}{2\pi}\frac{1}{t_{\text{orb},i}}\int_0^{2\pi} d\psi_\text{aps} \int_0^{t_\text{orb}} r_i(t)^2 dt \right)
	= \sum_i \frac{1}{2} m_i \omega_{\perp,i}^2 \,\left( \frac{1}{t_{\text{orb},i}} \int_0^{t_\text{orb}} r_i(t)^2 dt \right),
\end{equation}
where we assumed that the apsidal precession is so slow that $r(t)_i$ is independent from $\psi_\text{aps}$. The $r_i(t)$ describes the Keplerian ellipse Eq. (\ref{eq:r_K}) and we may use the results from Appendix \ref{app:Keplerian}. We have by use of equations (\ref{eq:E_K_dr}), (\ref{eq:r_pa_K}) that
\begin{equation}
	K_\text{VR} = \sum_i \frac{1}{2} m_i \omega_{\perp,i}^2 \,\frac{1}{2\pi}\sqrt{\frac{GM_i}{a_i^3}} 2\int_{r_{p,i}}^{r_{a,i}}\frac{r_i^2}{\dot{r}_i} dr_i 
	= \sum_i \frac{1}{2}m_i \omega_{\perp,i}^2 \,\frac{1}{\pi a_i}\int_{0}^{2e_i a_i}\frac{(r_i+r_{p,i})^3}{\sqrt{r_i(2e_ia_i - r_i)}} dr_i .
\end{equation}
Using the integrals (\ref{eq:integral_n}) it is straightforward to get
\begin{equation}\label{eq:K_I}
	K_\text{VR} = \sum_i \frac{1}{2} I_i \omega_{\perp,i}^2,
\end{equation}
where 
\begin{equation}\label{eq:I_inertia_app}
	I_i = \frac{1}{2}m_ia_i^2\left(1 + \frac{3}{2}e_i^2 \right).
\end{equation}
This is the moment of inertia of an annular disk $r_p\leq r\leq r_a$, about any of its diameters, with surface density
\begin{equation}
	\sigma(r) = \frac{1}{2\pi^2 a}\frac{1}{\sqrt{(r_a-r)(r-r_p)}}.
\end{equation}

The calculation of the planar term (spin) is straightforward. It equals the sum of mean kinetic energy of Keplerian orbits
\begin{equation}\label{eq:K_planar-bin_app}
	K_\parallel = \sum_i \frac{Gm_iM_i}{2a_i} = \text{const.}
\end{equation}
If we interpret this energy as the sum of spinning energies of the effective annular disks calculated above, we have
\begin{equation}\label{eq:K_planar_app}
	K_\parallel = \sum_i I_i \omega_{\parallel,i}^2,
\end{equation}
where the constant effective angular velocity is given by
\begin{equation}\label{eq:om_par_app}
	\omega_{\parallel,i} = \sqrt{\frac{GM_i}{2a_i^3}} \left( 1 + \frac{3}{2}e_i^3 \right)^{-1/2}.
\end{equation}

Using (\ref{eq:K_planar-bin_app}), (\ref{eq:E_Kepler}), the binary term (\ref{eq:U_B_def}) is straightforwardly given by
\begin{equation}
	U_\text{s} = -\sum_i \frac{Gm_iM_i}{a_i} = \text{const.}
\end{equation}

\bibliography{2019_Roupas_rigid_VRR}
\bibliographystyle{unsrt}

\end{document}